\documentclass[aps,floatfix]{revtex4}
\usepackage{epsfig}
\usepackage{color}
\usepackage{amssymb}
\usepackage{amsmath}
\usepackage[ngerman,english]{babel}
\usepackage{multirow}
\usepackage{graphicx}
\usepackage{subfigure}

\begin{document} 

\title{
%Non-abundant elements in the nuclear matter EOS\\
%Weakly-bound nuclei in the  nuclear matter EOS\\
%Leightest $p$-shell nuclei in the  nuclear matter EOS\\
Light  $p$-shell nuclei with cluster structures ($4 \le A \le 16$) in nuclear matter}
%Exotic, light  $p$-shell nuclei ($4 \le A \le 16$) in nuclear matter}

\author{Gerd R\"opke$^{1,2}$}
\affiliation{
$^1$Institut f\"ur Physik, Universit\"at Rostock, D-18051 Rostock, Germany\\
$^2$National Research Nuclear University (MEPhI), 115409 Moscow, Russia
}
\date{\today}

\begin{abstract}
The composition of hot and dense nuclear matter is calculated including the $1p$-shell nuclei $4 \le A \le 16$.
In-medium shifts, in particular Pauli blocking,  are determined by the intrinsic wave function of the nuclei. Results are given within a shell-model approach for the nucleon wave function.
Light  nuclei are not always well described by the shell model. The 'clustered' nucleus $^8$Be 
exhibits strong correlation effects because of $\alpha$-like clustering. Intrinsic cluster structures are also significant for the nuclei $^6$Li, $^7$Li, $^7$Be, and $^9$Be.
The contribution of the relatively rare %non-abundant 
elements Li, Be, and B, to the equation of state (EoS) of matter near the saturation density is overestimated in simple approaches such as the nuclear statistical equilibrium (NSE) model. 
Both, the treatment of continuum correlations and the account of in-medium modifications are considered for the contribution of $^5$He and $^4$H clusters. Compared to the extended NSE including unstable nuclei,
the contributions of the corresponding P$_{3/2}$ channel with $A=5, Z=2$ and  P$_{2}$ channel with $A=4, Z=1$, respectively, to the EoS are strongly suppressed at high densities owing to Pauli blocking effects. For the shifts of the binding energies of the light $p$-shell nuclei, simple fit formula are given to calculate the composition of hot and dense matter in a wide parameter range.
\end{abstract}


\maketitle

\section{Introduction}

Nuclear systems, such as nuclei, excited matter produced in heavy ion collisions, 
as well as nuclear matter which is found in compact astrophysical objects, are strongly coupled quantum systems.
The traditional treatment \cite{RS} of dense nuclear systems which is based on a single-nucleon quasiparticle approach, 
such as the relativistic mean-field approximation, the shell model of nuclei, or the transport models related 
to the Boltzmann equation, has to be improved to describe quantum correlations,
in particular the formation of bound states (clusters). Four nucleon, $\alpha$-like correlations
have been considered to describe nuclei such as the Hoyle state of $^{12}$C, see \cite{THSR,THSR2} and references given there,
but are also of relevance to describe the $\alpha$ decay of heavy nuclei \cite{Xu}.
The production of clusters in heavy-ion collisions (HIC), see, e.g., \cite{Armstrong,Ganil}, demands the treatment of clusters in highly excited nuclear matter
what can be  realized within a quantum statistical approach \cite{RMS82}. In thermodynamic equilibrium, in
simplest approximation a mass action law is obtained describing chemical equilibrium in a mixture of ideal, non-interacting
components performing reactive collisions, which is denoted as nuclear statistical equilibrium (NSE) \cite{NSE}. 
Improvements are obtained taking into account excited states, in particular 
the contribution of the continuum to obtain virial expansions \cite{RMS82,BU,HS}.  
In non-equilibrium, codes such as the antisymmetrized molecular dynamics (AMD) and quantum molecular dynamics (QMD) 
have been developed to include cluster formation in the treatment of HIC, see Ref.  \cite{Wolter}.
The equation of state of stellar matter in a wide range of temperature $T$, baryon density $n_B=n_n^{\rm tot}+n_p^{\rm tot}$, and asymmetry $Y_p=n_p^{\rm tot}/n_B$ is of interest in supernovae explosions, see \cite{Fischer,Fischerarx} and references given there,
and the account of few-nucleon correlations and cluster formation is relevant for the treatment of different processes
during the evolution of compact astrophysical objects. Alternatively to  the total 
number densities $n^{\rm tot}_\tau$ of neutrons ($\tau = n$) 
and protons ($\tau = p$),
the state of nuclear matter can also be described by  the chemical potentials $\mu_\tau$,  in addition to $T$. 

The simple NSE and its improvements considering excited states and scattering phase shifts of the isolated few-nucleon 
problem cannot be applied to baryon number densities near the saturation density $n_{\rm sat}=0.15$ fm$^{-3}$ where the interaction between 
the constituents cannot be neglected. A systematic quantum statistical approach to thermodynamic equilibrium can be given which
uses the concepts of Green's functions, spectral functions, and frequency-dependent self-energy, for which a cluster decomposition
can be performed.
A main feature is that  bound states can be treated as quasiparticles with medium dependent binding energies and wave functions. 
They are obtained from  an in-medium Schr\"odinger equation derived within a Green's function approach \cite{RMS82}. 
In addition to the single-nucleon self-energy, the antisymmetrization of the wave function (Pauli principle) is of relevance. 
Starting from the mass action law at low densities, clusters become less bound at increasing densities 
because of Pauli blocking. They are dissolved at a critical density so that near the saturation density 
a Fermi liquid model of single-nucleon quasiparticles becomes applicable. 
In particular, the contribution of two-nucleon correlation has been discussed at arbitrary densities 
\cite{SRS}. The inclusion of light clusters $A \le 4$, i.e. deuteron $d$ ($^2$H), triton $t$ ($^3$H), helion $h$ ($^3$He), and  $\alpha$ ($^4$He),  has also been investigated, see \cite{r3} and references given there.
Only first steps have been made to  include higher clusters $A > 4$ \cite{Debrecen} within this approach.

The present work is devoted to the investigation of clusters with mass number $4 \le A \le 16$ 
where, in addition to the $1s$ shell, the $1p$ shell is filled. A list of the corresponding stable nuclei is given in Tab. \ref{Tab:1},
together with some known properties.
There are some recent works to include $1p$-shell nuclei in the calculation of the equation of state and the composition of hot and dense matter.
In \cite{Yudin}, unstable, neutron-rich isotopes such as $^4$H, $^5$He, and isotopes with even higher neutron content have been included in the NSE.
A strong dominance of neutron-rich isotopes is found at high densities and low proton fraction $Y_p$. 
In-medium effects may be included within an excluded-volume approach \cite{Hempel} but 
the dominance of unstable, neutron-rich isotopes at high densities remains. Another approach to include $1p$ nuclei in the EoS \cite{Pais}
has been proposed within a generalized RMF approach \cite{TRB} where all nuclei are considered as new quasiparticles, and the corresponding fields
are coupled to the meson fields. These semi-empirical approaches should be founded by a more systematic quantum statistical 
approach as indicated in this work.

We focus on two aspects of the inclusion of light $p$-shell nuclei, the in-medium modification and dissolution of bound states at increasing density owing to Pauli blocking 
and the account of continuum correlation within a generalized cluster Beth-Uhlenbeck approach. We propose fit formulas to reproduce the 
energy shifts and the virial coefficients, i.e. the partial intrinsic partition functions, which are needed to calculate the composition of nuclear matter
in a wide parameter range. These expressions can be used for the evaluation of the EoS, but are also of interest for other
applications such as kinetic and transport processes in sub-saturation nuclear matter, see \cite{Wolter} and references given there.

The paper is organized as follows: After a short review of the formalism in Sec. \ref{Sec:formalism} with the focus on Pauli blocking,
we discuss in Sec. \ref{Sec:boundshift} the in-medium shifts of the binding energy of bound $1p$ nuclei. Because Pauli blocking is connected with the occupation 
in phase space, the wave function of the bound states in momentum representation is essential and will be discussed in Sec. \ref{Sec:wavefct}.
To discuss the contribution of unstable nuclei such as $^4$H, $^5$He, it is necessary to consider the continuum correlations in Sec. \ref{Sec:virial}.
Exemplary calculations for the composition of nuclear systems are presented and discussed in Sec. \ref{Sec:calculations}. We find that in comparison to the NSE,
the mass fraction of $1p$ nuclei is significantly reduced near the saturation density if in-medium effects are systematically taken into account.

\section{Basic expressions}
\label{Sec:formalism}

\subsection{Composition of dense nuclear matter}

We employ a strict quantum statistical approach to nuclear matter in thermodynamic equilibrium, 
characterized by the temperature $T$ and the chemical potentials $\mu_\tau$ \cite{r3}. Neglecting weak processes, there are two conserved quantities, 
the total number of neutrons  and protons (bound in nuclei and free ones) with the corresponding chemical potentials $\mu_n, \mu_p$. 
As an equation of state, the total densities $n^{\rm tot}_\tau$ of neutrons 
and protons are obtained using the method of thermodynamic Green's functions via the single-nucleon 
spectral functions or the related self-energy.
At subsaturation baryon densities $n_B=n^{\rm tot}_n+n^{\rm tot}_p \le n_{\rm sat}$, 
we are interested in cluster formation which is described by the cluster decomposition of the 
self-energy. As a result, the total densities of neutrons/protons are given as the sum of 
free nucleons and the nucleons bound in clusters,
\begin{eqnarray}
\label{eos}
&&  n^{\rm tot}_n (T,\mu_n,\mu_p)=  \frac{1 }{ \Omega} \sum_{A,Z,J,\nu,{\bf P}}N 
f_{A,Z}\left(E_{A,Z,J,\nu}({\bf P};T,\mu_n,\mu_p)\right) =\sum_{A,Z,J}N 
n^{\rm part}_{A,Z,J}(T,\mu_n,\mu_p),  \nonumber\\ 
&&  n^{\rm tot}_p(T,\mu_n,\mu_p)= \frac{1 }{ \Omega} \sum_{A,Z,J,\nu,{\bf P}}Z 
f_{A,Z}\left(E_{A,Z,J,\nu}({\bf P};T,\mu_n,\mu_p)\right)=\sum_{A,Z,J}Z 
n^{\rm part}_{A,Z,J}(T,\mu_n,\mu_p)  \, ,
\label{quasigas}
\end{eqnarray}
i.e., the sum over the partial densities of the different channels characterized by $\{A,Z,J\}$. 
$N=A-Z$ is the neutron number, $\Omega$ the volume, and
$\bf P$ denotes the center-of-mass (c.m.) momentum of the cluster (or, for $A=1$, the momentum of the nucleon). 
The internal quantum state $\nu$ describes possible intrinsic excitations of the $A$-nucleon cluster, and
\begin{equation}
f_{A,Z}(\omega;T,\mu_n,\mu_p)=\frac{1}{ \exp [(\omega - N \mu_n - Z \mu_p)/T]- (-1)^A}
\label{vert}
\end{equation}
is the Bose or Fermi distribution function for even or odd $A$,
respectively. 
We are interested in parameter values where the free nucleons may become degenerate. 
For all other clusters the classical approximation is possible at $T >1$ MeV.

In the low-density, low temperature limit we take the ground-state energies (the negative of the binding energies)
\begin{equation}
\label{E0}
E_{A,Z,J,\nu}({\bf P};T,\mu_n,\mu_p)\approx E^{(0)}_{A,Z,J}+\hbar^2P^2/(2 A m) 
\end{equation} 
and perform the summation over $\bf P$ and $\nu$ (degeneracy factor $2 J+1$) so that the partial density of channel $\{A,Z,J\}$ results as
\begin{equation}
\label{zpart0}
 n^{\rm part,0}_{A,Z,J}(T,\mu_n,\mu_p)=(2J+1) \left(\frac{Am T}{2 \pi \hbar^2}\right)^{3/2} 
e^{\left(-E^{(0)}_{A,Z,J}+N\mu_n+Z\mu_p\right)/T}.
\end{equation}
Here, $m$ denotes the nucleon mass (we neglect the proton - neutron mass difference). 
The bound state energies $E^{(0)}_{A,Z,J}=-B_{A,Z}$ and the degeneracy $2J+1$ are found in the tables of nuclei \cite{AudiWapstra,nuclei}.
This approximation for the EoS is also denoted as nuclear statistical equilibrium (NSE).
It describes an ideal mixture of nuclei (bound states), interacting only occasionally via reactive collisions. 

The  simple approximation (\ref{zpart0}) can be improved in different ways. First, 
not only the ground state $E^{(0)}_{A,Z,J}$, but also the excited states $\nu$ of the nucleus 
with quantum numbers $\{A,Z,J\}$ contribute to the partial densities (\ref{zpart0}). 
In particular, the scattering states describing continuum correlations have to be taken into account. 
If the scattering states of two clusters are described by the scattering phase shifts $\delta_{A,Z,J}(E)$
with $\nu \to E$ as the energy of relative motion, the virial EoS is derived from a quantum statistical 
approach \cite{BU,RMS82,SRS,HS,clustervirial,VT}.
We discuss this contribution of scattering states as given by the Beth-Uhlenbeck equation in Sec. \ref{Sec:virial}.

Secondly, with increasing density, the approximation of non-interacting clusters is no longer possible, 
and medium modifications have to be considered. A quantum statistical approach can be used, 
see \cite{r3} and further references given there. In particular, a quasiparticle approach can be given
where the energies of the nucleons and of the nuclei,  $E_{A,Z,J,\nu}({\bf P};T,\mu_n,\mu_p)$, 
are depending on the temperature and baryon densities of the nuclear medium. 
In addition, the dependence on the c.m. momentum ${\bf P}$ is more general than the expression (\ref{E0}).
These modifications are given by the self-energy of the single-nucleon states and the Pauli blocking 
on the interaction within the clusters, for details see \cite{r3} for $A \le 4$. 
Also the bound-state wave functions and the scattering phase shifts are modified. 
We discuss these medium modifications for the bound states with $4 \le A \le 16$ in Sec. \ref{Sec:boundshift}.

\subsection{In-medium shift of bound nuclei}
\label{Sec:Pauli}

In the low-density limit, the virial form of the EoS can be calculated knowing the empirical values of 
the cluster binding energies and the scattering phase shifts. The knowledge of the interaction potential is not necessary. 
This is not the case at higher densities where the medium modifications have to be taken into account. 
Within the quantum statistical approach,
% we have to solve the propagator
%of a few-body cluster using, e.g., the Green-function method. In particular, 
we have to solve the $A$-particle in-medium 
Schr{\"o}dinger equation (momentum representation)
\begin{eqnarray}
\label{Awave}
 &&[E_{\tau_1}^{\rm qu}(1)+\dots +E_{\tau_A}^{\rm qu}(A)-E_{A,Z,J,\nu}({\bf P};T,\mu_n,\mu_p)]\Psi_{A \nu P}(1\dots A) \nonumber \\
&&+\sum_{1'\dots A'} \sum_{i  < j}[1-f_{\tau_i}(i)-f_{\tau_j}(j)] V(ij,i'j') \prod_{k\neq i,j} \delta_{kk'}\Psi_{A \nu P}(1'\dots A')=0,
\end{eqnarray}
where $1=\{{\bf p}_1,\sigma_1,\tau_1 \}$ denotes momentum, spin, and isospin variables. 
$E_{\tau_1}^{\rm qu}(1)$ are quasiparticle energies which are obtained from a frequency-dependent self-energy. 
We can use parametrizations \cite{TRB} derived from relativistic mean-field approximations such as  DD2-RMF \cite{Typel} or an effective mass approximation. 
The self-energy shift acts for the bound states as well as for the continuum and has no influence
on the binding energy in the rigid shift approximation where the $\bf p$-dependence of the shift is neglected. 
Then, it can be implemented in the chemical potential.
Within the  effective mass approximation, a minor effect on the shift of the binding energy was obtained in \cite{r1}.

\begin{figure}[h]
\centerline{\includegraphics[width=120pt,angle=0]{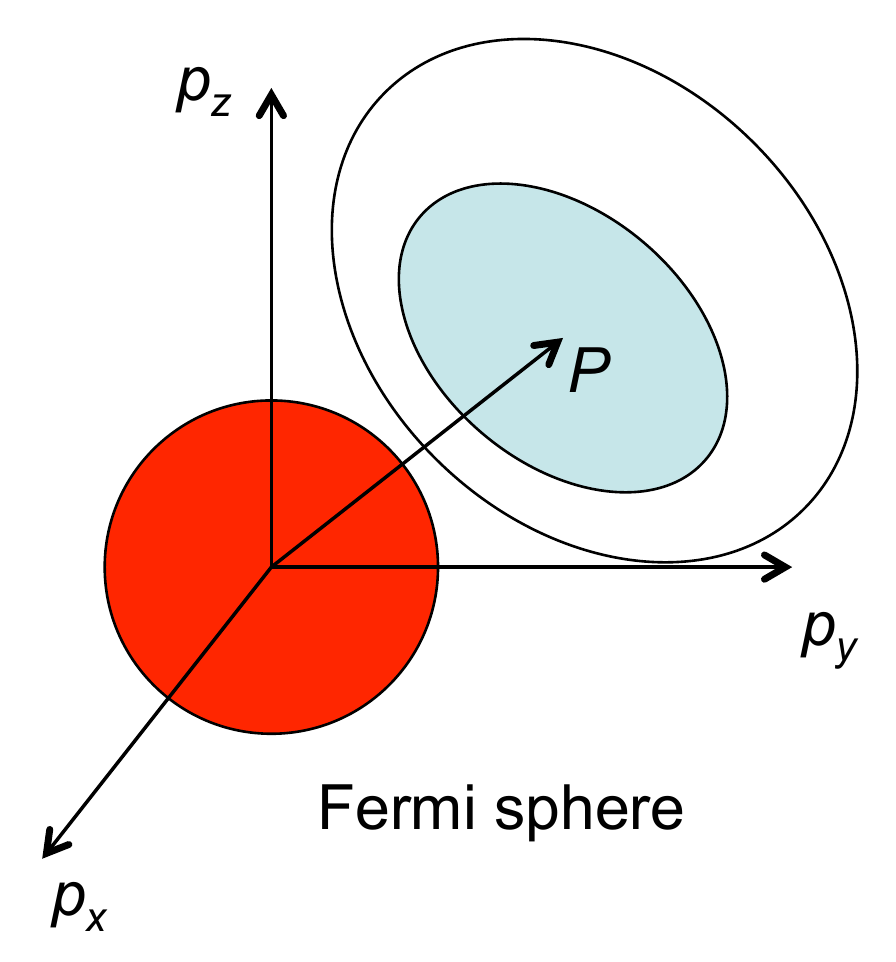}}
\caption{Pauli blocking: In momentum space $\{p_x,p_y,p_z\}$, the Fermi sphere is occupied. A bound state is formed around the c.m. momentum $\bf P$ using free phase space.
Contributions of the occupied Fermi sphere cannot be used to form the bound state wave function.}
\label{fig:P}
\end{figure}

More important is the Pauli blocking given by the occupation number $f_{\tau_i}(i)$ of the single-nucleon state $i$ in front of 
the interaction potential in Eq. (\ref{Awave}). Neglecting the correlations in the surrounding medium, 
we can use a Fermi distribution function with effective values for temperature and chemical potential to approximate the 
actual occupation numbers. 
Single-nucleon states which are already occupied cannot be used to built up the bound state wave function 
$\Psi_{A \nu P}(1\dots A)$. As a consequence, the binding energy $-E_{A,Z,J,\nu}({\bf P};T,\mu_n,\mu_p)$ is shifted, depending on the 
cluster intrinsic quantum numbers $\nu$ and the c.m. momentum $\bf P$, 
but also on temperature $T$ and chemical potentials $\mu_\tau$. A schematic representation of the Pauli blocking and its dependence on 
the c.m. momentum $\bf P$ is shown in Fig. \ref{fig:P}.

For the light clusters $A \le 4$, the Pauli blocking has been discussed in the literature \cite{r1,r2,r3}.
An effective separable potential  $V(ij,i'j')$ has been considered which reproduces known properties, 
in particular the binding energies and rms radii.
Similar calculations to determine the in-medium energy shifts as function of temperature, densities of protons/neutrons,
and the c.m. momentum can also be performed for larger, weakly bound clusters, see Appendix \ref{App:6Li}. 

 For large numbers $A$, an appropriate description of the nucleon wave function is the shell model where 
 the $A$-nucleon wave function is approximated by the antisymmetrized product (Slater determinant) of single-nucleon
  wave functions obtained from an effective potential $V^{\rm mf}(1,1')$. A widely used local potential is the 
Woods - Saxon potential \cite{WoodsS}. Bound and scattering states are easily obtained from a separable potential \cite{Yama,Mongan}
which model will also be used here, see Appendix \ref{Sec:separabel}. Note that any local potential can be expressed as sum of separable potentials \cite{EST}.

Within the shell model, the nucleons are moving independently on single-particle orbits. 
Instead of Eq. (\ref{Awave}), we have to solve the single-nucleon wave equations
\begin{eqnarray}
\label{1wave}
 &&E_{\tau_1}^{\rm qu}(1)\,\psi_{1 \nu P}(1) %\nonumber \\&&
+\sum_{1'} [1-f_{\tau_1}(1)]\, V^{\rm mf}(1,1')\,\psi_{1 \nu P}(1')=E_{1\nu}^{\rm qu}({\bf P};T,\mu_n,\mu_p)\,\psi_{1 \nu P}(1),
\end{eqnarray}
where the dependence of the c.m.~momentum $\bf P$ results from the relative motion of the Fermi distribution $ f_{\tau_1}(1) $. 
The Pauli blocking shift of the single-nucleon states follows as 
\begin{equation}
\label{paulnu}
 \Delta E^{\rm Pauli}_{1\nu}({\bf P};T,\mu_n,\mu_p)=\sum_{11'}\psi_{1 \nu P}(1)\,f_{\tau_1}(1)\,V^{\rm mf}(1,1')\,\psi_{1 \nu P}(1').
\end{equation}

An important ingredient to calculate in-medium effects is the nucleon wave function of the $A$-nucleon cluster. In the simplest form of a density functional approach, the Thomas-Fermi model, the 
many-particle wave function is approximated locally by plane waves, and shell effects are not described. The shell model starts
from the approximation of the antisymmetrized product of single nucleon quasiparticle orbits and has to include correlation effects.
Alternative concepts to approximate the many-nucleon wave function are based on the cluster model which adequately includes,
for instance, $\alpha$-like clustering in light nuclei, in particular $^8$Be or the Hoyle state of $^{12}$C, see \cite{THSR,THSR2} and Sec. \ref{sec:clustering} below.

\section{Pauli blocking of $p$-shell nuclei}
\label{Sec:boundshift}

\subsection{Intrinsic nucleon wave function of a cluster}
\label{Sec:wavefct}

To calculate the in-medium shifts, the intrinsic wave function of the nucleons in the nucleus (cluster) is needed. 
We focus here on the Pauli blocking which is responsible for the disappearance of bound states with increasing density. 
As seen from Fig. \ref{fig:P}, this effect is determined by the wave function in momentum space and the overlap with the Fermi distribution function. 
Therefore, in this section  we try to find appropriate approximations for the intrinsic wave function.
The self-energy corrections  cancel nearly with the shift of the continuum and give only a small contribution to the in-medium shift of the binding energy which describes the energy difference,
but must be included in the bound state energy $E_{A,Z,J,\nu}$, see \cite{r1} and Sec. \ref{Sec:calculations}.

We try to extract the wave function from empirical data, in particular the rms radii, cf. \cite{r2} for the light $1s$ nuclei $A \le 4$. 
In the following  section \ref{sec:shell}, we consider the nuclear shell model. 
The $1p$ nuclei with $5\le A\le 16$ are described by the successive occupation of the $1p$ orbit. 
We consider independent mean-field orbitals, correlations and spin-orbit interaction are neglected.
To treat strong correlations in the nucleon wave function, the formation of subclusters is discussed in Sec. \ref{sec:clustering}.

We use Gaussian wave functions which have the advantage that the center-of-mass (c.m.) motion can be separated from the intrinsic motion.
The Gaussian wave function has been considered in \cite{r1} for the light nuclei and compared to a Jastrow function approach. 
The differences of the results for the Pauli blocking shift are small so that
we conclude that details of the wave function are not very important, only the global distribution in phase space and the overlap with the Fermi sphere is relevant. 

The shell-model wave functions of interest are the $1s$ and $1p$ states with different width parameter $B_s, \,B_p$, respectively,
\begin{equation}
\psi_{1s}({\bf r}) \propto e^{-r^2B_s^2/4}\,Y_{00}(\theta,\phi), \qquad \psi_{1p}({\bf r}) \propto e^{-r^2B_p^2/4}\,r\,Y_{1m}(\theta,\phi)
\end{equation}
or, in Fourier space,
\begin{equation}
\psi_{1s}({\bf p}) \propto e^{-p^2/B_s^2}\,Y_{00}(\theta,\phi), \qquad \psi_{1p}( {\bf p}) \propto e^{-p^2/B_p^2}\,p\,Y_{1m}(\theta,\phi).
\end{equation}
The ratio of the squared width parameters will be denoted by $\beta=B_s^2/B_p^2$.
The $A$-nucleon wave function 
\begin{equation}
\label{psiA}
 \Psi_{A,\nu}(1,\dots,A)={\cal A}\{\psi_{1s,\nu_1}(1)\dots\psi_{1s,\nu_4}(4)\psi_{1p,\nu_5}(5)\dots \psi_{1p,\nu_A}(A)\}
\end{equation}
is approximated by the antisymmetrized product (Slater determinant) of  occupied orbitals, $\nu$ denotes the quantum numbers of the cluster, 
and the intrinsic quantum number $\nu_i$ contains spin and isospin of the single nucleon.

The point rms radius of the $A$-nucleon cluster follows as square root of
\begin{equation}
 \langle r^2 \rangle_{A,\nu} = \frac{1}{A}\langle \Psi_{A,\nu}|\sum_i^A({\bf r}_i-{\bf R}_{\rm cm})^2 |\Psi_{A,\nu}\rangle
\end{equation}
with the c.m. position ${\bf R}_{\rm cm}=A^{-1}\,\sum_i^A {\bf r}_i$. For $A \le 4$ the nucleons occupy $1s$ orbits. After introduction of Jacobi coordinates, 
the c.m. part can be separated, and the intrinsic part gives \cite{r1} 
\begin{equation}
\label{rms4}
 \langle r^2 \rangle_{A,\nu} = \frac{3 (A-1)}{A \, B_s^2}, \qquad A \le 4.
\end{equation}
The same result is obtained if we take ${\bf R}_{\rm cm}=0$ so that ${\bf r}_1=-{\bf r}_2-\dots-{\bf r}_A$.

For larger nuclei $4 \le A\le 16$ the $1p$ orbitals are successively occupied. If we assume $\beta = 1$, i.e., we assume  $B_s=B_p$ and denote this common value as $\bar B$, we obtain for the point rms radii the square root of
\begin{equation}
\label{rms16}
 \langle r^2 \rangle_{A,\nu} = \frac{3 (A-1)+2 (A-4)}{A \, {\bar B}^2}, \qquad 4\le A \le 16,
\end{equation}
which can be used to derive this common parameter  $\bar B$ from the observed rms radii.

For $\beta=B_s^2/B_p^2 \neq 1$, the expressions for the rms radii are more complex. As example, for $A=6$ we find (\ref{rmsLi})
\begin{equation}
\label{rms6beta}
\langle r^2 \rangle_{^6{\rm Li}}= \frac{\beta}{6 B_s^2} \frac{21 + 160/\beta + 382/\beta^2 + 688/\beta^3 + 288/\beta^4}{(1 + 
        2/\beta) (3 + 8/\beta + 16/\beta^2)}. 
\end{equation}

For $4 \le A \le 16$ and arbitrary $\beta$, as approximation to results like Eq. (\ref{rms6beta}) we assume the sum 
of a contribution from the $1s$ orbit (4 nucleons) and a contribution from the $1p$ orbit [$(A-4)$ nucleons],
\begin{equation}
\label{rmsbetafit}
 \langle r^2 \rangle_{A,\nu} \approx \frac{9}{A \, {B_s}^2}+\frac{5 A-20}{A \, {B_p}^2}, \qquad 4\le A \le 16.
\end{equation}
We demand that the nucleon wave functions should reproduce the measured rms radii shown in Tab. \ref{Tab:1}. This is essential for the correct determination of the distribution in momentum space and the calculation of Pauli blocking.

\begin{table}
\begin{center}
\hspace{0.5cm}
 \begin{tabular}{|c|c|c|c|c|c|c|c|c|c|}
\hline
$A$  &  $Z$ &  $\frac{B_{A,Z}}{A} $ [MeV]&  $ g_{A,Z} $ &abundance/half-life & rms$_{\rm  charge}$ [fm] 
& rms$_{\rm point}$ [fm] & ${\bar B}$ [fm$^{-1}$]&  $B_s$  [fm$^{-1}$] & $\beta$\\
\hline
1& 1 & - & 2 & 12 [0.99998] & 0.8783 & 0 & -& - & -\\
2& 1 & 1.112 & 3 &12 [0.00002] & 2.1421 & 1.9538 & 0.627& - & -\\
3& 1 & 2.827 & 2 & 12.32 y ($\beta^-$) & 1.7591 & 1.5242 &0.928 & - & -\\
3& 2 & 2.572 & 2 &  10.93 [0.000166] & 1.9661 & 1.7590 & 0.804& - & -\\
4& 2 &  7.073 & 1 & 10.93 [0.999834]  & 1.6755 & 1.427& 1.051& 1.051 & -\\
5& 2 & 5.512 & 4 &  $2.04 \times 10^{-22}$ s & - & - & -& - & -\\
6& 3 & 5.332 & 3& 1.05 [0.07594]  & 2.589 & 2.435 & 0.731& 0.982 & 2.533\\
7& 3 & 5.606 & 4   & 1.05 [0.9241] & 2.444 & 2.281 & 0.812& 0.957 & 1.626\\
7 & 4 & 5.372 & 4 & 53 d (ec) & 2.646 & 2.496 & 0.742& 0.957 & 2.065\\
8 & 4 & 7.062 & 1 &  $8.19 \times 10^{-17}$ s & - & - & -& - & -\\
9 & 4 & 6.462 & 4  & 1.38 [1.0] & 2.519 & 2.438 & 0.797& 0.918 & 1.444\\
10 & 4 & 6.497 & 1  & 1.5 Gy ($\beta^-$) & 2.355 & 2.185 & 0.904& 0.902 & 0.995\\
11 & 4 & 5.953 & 2  & 13 s ($\beta^-$) & 2.463  & 2.301 & 0.869& 0.888 & 1.055\\
10 & 5 & 6.475 & 7   & 2.70 [0.199] & 2.4277 & 2.263 & 0.873& 0.902 & 1.089\\
11 & 5 & 6.928 & 4  & 2.70 [0.801]  & 2.4060 & 2.240 & 0.893& 0.888 & 0.986\\
12 & 6 & 7.680 & 1 & 8.43 [0.98894] & 2.4702 & 2.309 & 0.875& 0.875 & 1.0\\
13 & 6 & 7.470 & 2 & 8.43 [0.01062] & 2.4614 & 2.2994 & 0.886& 0.864 & 0.939\\
14 & 6 & 7.520 & 1 & 5700 y ($\beta^-$) & 2.5025 & 2.3433 & 0.876& 0.853 & 0.938\\
14 & 7 & 7.476 & 1 & 7.83 [0.99771] & 2.5582 & 2.4027 & 0.854& 0.853 & 0.996\\
15 & 7 & 7.699 & 2 & 7.83 [0.00229] & 2.6058 & 2.4533 & 0.842& 0.843 & 1.003\\
16 & 8 & 7.976 & 1 & 8.69 [0.99762] & 2.6991 & 2.5522 & 0.814& 0.834 & 1.059\\
\hline
 \end{tabular}
\caption{Data of  (nearly) stable nuclei  $A \le 16$ as well as $^5$He, $^8$Be. Mass number  $A$, charge number $Z$, binding energy per nucleon $B_{A,Z}/A $, degeneracy factor $g= 2J+1$ \cite{nuclei}. Solar element abundance: $^{10}$log relative to 12 for Hydrogen, [square brackets]:  isotope fraction \cite{solar}. Half-life in s(seconds), d(ays), G(iga)y(ears) according to Ref. \cite{nuclei}. Charge rms radii taken from Ref.  \cite{rmsradii}. Parameter values $\bar B$ (\ref{rms16}), the ansatz 
$B_s(A) = 1.324\,A^{-1/6}$ fm$^{-1}$ as well as the parameter  $\beta=B_s^2/B_p^2$ are also given.}
\label{Tab:1}
\end{center}
\end{table}

In Tab. \ref{Tab:1}, the  stable nuclei with mass number $A \le 16$ are shown, together with  nuclei with half-life larger than 1 s. For comparison, the nucleus $^5$He and the interesting nucleus $^8$Be are also included.
The binding energy per nucleon $B_{A,Z}/A$ \cite{nuclei} and degeneracy are given. Note that the binding energy per nucleon for $^8$Be is quite large compared to the 
neighboring nuclei. However, it is not stable as shown by the very short half-life. It decays into  two $\alpha$ particles which have 
even higher values for the binding energy per nucleon. 
There is also no stable nucleus with the mass number $A=5$. Within the shell model approach, 
the nucleon added to the $^4$He core has to be positioned in the $1p$ state at higher kinetic energy so that binding does not occur. 
We discuss $^5$He-like correlations below in Sec. \ref{Sec:5He}. 
All other nuclei have a binding energy larger than the sum of the binding energies of respective cluster components.

The unstable, long-living isotopes given in Tab. \ref{Tab:1} have weak interaction decays (electron capture for $^7$Be, $\beta^-$ for $^3$H, $^{10}$Be, and $^{11}$Be).
For the stable nuclei, the solar element abundances are given, in addition also the isotope fractions  \cite{solar}.  
 Compared to the elements C, N, O, the clustered 'rare' elements Li, Be, B have a very low abundance. 
Note that missing bound nuclei with $A=5,8$ are relevant for nucleosynthesis in astrophysics.

In Tab. \ref{Tab:1}, values for the charge rms radii and point rms radii, ${\rm rms}_{\rm point}^2={\rm rms}_{\rm charge}^2-0.8783^2\, {\rm fm}^2$, are taken from Ref. \cite{rmsradii}.
The rms radii do not exhibit a simple dependence on $A$ as expected, e.g., for a liquid drop model.
Details of the $A$-nucleon wave function are of relevance. The nuclear shell model describes already important properties 
of the $A$-nucleon wave function. Correlations which also influence the rms radii, in particular clustering \cite{THSR}, are discussed below in Sec. \ref{sec:clustering}.

\begin{figure}[h]
\centerline{\includegraphics[width=350pt,angle=0]{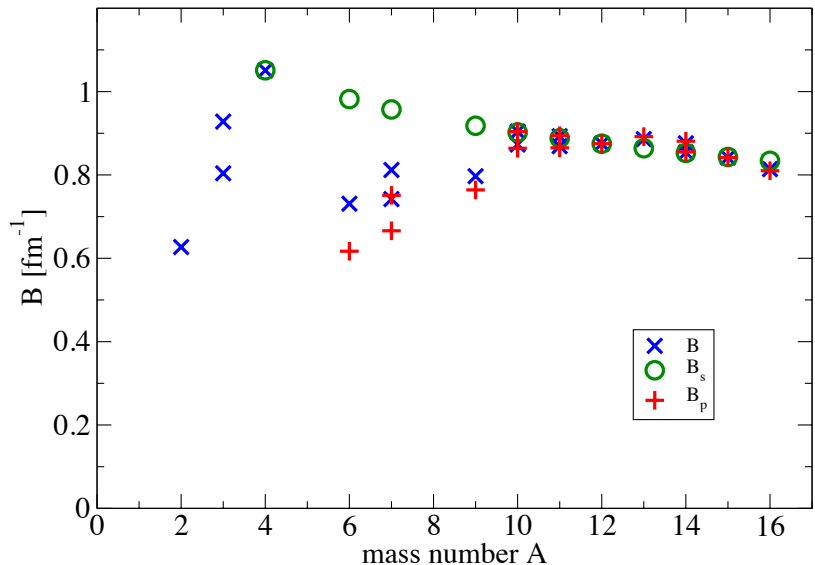}}
\caption{The range parameter $B$ of the Gaussian orbits. $\bar B$ according (\ref{rms16}) (blue crosses)
as well as $B_s$ (green circles) and $B_p$ (red plus) as function of the mass number $A$.}
\label{fig:0}
\end{figure}

The deuteron is weakly bound and, therefore, extended in configuration space. The difference of the rms radii of $t$ and $h$
is well understood, see \cite{r2}, App. A. The $\alpha$ particle is a compact, strongly bound nucleus. 
The wave functions of these light nuclei are reasonably described by a Gaussian $1s$ wave function \cite{r1}. The calculation of $\bar B$ according to Eq. (\ref{rms16}) assuming $\beta=1$ 
is shown in Tab.  \ref{Tab:1}, see also Fig. \ref{fig:0}. 
A smooth behavior is obtained for $A  \ge  10$. 
The clustered nuclei with $6 \le A \le 9$ demand a further discussion of the nucleon wave function.

%\begin{figure}[h]
%\centerline{\includegraphics[width=350pt,angle=0]{fitB.pdf}}
%\caption{The range parameter $B$ of the Gaussian orbits. $\bar B$ according (\ref{rms16}) (blue crosses)
%as well as $B_s$ (green circles) and $B_p$ (red plus) as function of the mass number $A$.}
%\label{fig:0}
%\end{figure}

In the general systematics, see \cite{rmsradii}, the rms value for $^6$Li is relatively large. 
Within the nuclear shell model approach, two nucleons are positioned in the $1p$ state weakly bound to the $\alpha$-like core.
We can account for weakly bound, more extended nucleons in the $1p$ state 
if we construct the shell model wave function (\ref{psiA}) with different parameter values $B_s,B_p$, i.e. with $\beta \neq 1$.
Strong deviations are expected for the clustered nuclei $^6$Li,  $^7$Li, $^7$Be, and $^9$Be, whereas the nuclei with $A \ge 10$ behave smoothly.

For an exploratory calculation within a shell model approach, we assume that the inner $1s$ wave function 
changes smoothly if $A$ is increasing. With the ansatz $B_s(A) = 1.324\,A^{-1/6}$ fm$^{-1}$
the values $\bar B,$ Tab. \ref{Tab:1}, are approximately reproduced for $A \ge 10$. Then, the parameter values $B_p$ given in Tab. \ref{Tab:1} follow from Eq. (\ref{rmsbetafit}).
For $A=6$, the value $B_s(6) = 0.982$ fm$^{-1}$ is estimated. To reproduce the empirical value of the point rms radius of $^6$Li,
with Eq. (\ref{rms6beta}) we find $\beta = 2.533$, i.e. $B_p = 0.617$ fm$^{-1}$. 
These values are also confirmed by a more detailed six-nucleon calculation given below, Appendix \ref{App:6Li}. 

The results shown in Tab. \ref{Tab:1} and Fig. \ref{fig:0} describe only properties of the wave function as derived from the rms radii. 
The $\alpha$-like core changes smoothly, but the outer $1p$ nucleons show a particular  behavior for $5 \le A \le 9$. 
Small values of $B_p$ means that the $1p$ orbital is very extended. As a consequence, the density is low, and correlation effects  become relevant.
A signature is cluster formation which appears in the low-density regions, as known from the Hoyle state. Here, the many-nucleon wave function has another structure. 
For $^8$Be, it is
described in good approximation by the THSR wave function \cite{THSR}. Clustering in nuclei \cite{Lyu,Zhao,Kanada} will be discussed below in Sec. \ref{sec:clustering}.

\subsection{Shell-model approach}
\label{sec:shell}

\begin{table}
\begin{center}
\hspace{0.5cm}
 \begin{tabular}{|c|c|c|c|c|c|c|c|c|c|}
\hline
$A$  &  $Z$ &  $B_s$ &  $V^{\rm WS}_{0,s} $& $B_p$  &  $V^{\rm WS}_{0,p} $
& $a_{AZ}$ &$b_{AZ}$ & $f_{AZ}$& $ g_{AZ} $\\
{} & {} & [fm$^{-1}$] & [MeV] & [fm$^{-1}$] & [MeV] &[MeV fm$^3$]& [MeV$^{-1}$]& [MeV$^{5/2}$ fm$^3$]& [MeV]\\
\hline
4  & 2 &  1.051 & 73.7 &  -      & -    		 	& 796.1& 0.06002 & 50621  &  14.291\\
6  & 3 &  0.982 & 63.8 &  0.617  & 35.2 	 	& 640.6& 0.06427 & 35278  &  12.771\\
7  & 3 &  0.957 & 60.8 &  0.751 & 41.6 	 	& 599.9& 0.06188 & 35845  &  13.624\\
7  & 4 &  0.957 & 60.8 &  0.666  & 35.2 	 	& 598.4& 0.06440 & 32834  &  12.737\\
9  & 4 &  0.918 & 56.8 &  0.764  & 40.5 	 	& 549.3& 0.06094 & 33943  &  13.990\\
10 & 4 &  0.902 & 55.3 &  0.904  & 61.0   		& 541.5& 0.05148 & 47499  &  18.281\\
11 & 4 &  0.888 & 54.3 &  0.865 & 54.1 	 	& 532.4& 0.05290 & 46823  &  18.482\\
10 & 5 &  0.902 & 55.3 &  0.864 & 53.8 	 	& 539.4& 0.05414 & 42678  &  16.916\\
11 & 5 &  0.888 & 54.3 &  0.894 & 59.7	 	 & 534.8& 0.05068 & 48450  &  18.730\\
12 & 6 &  0.875 & 53.3 &  0.875  & 56.6 	 	& 529.1& 0.05085 & 47663  &  18.653\\
13 & 6 &  0.864 & 52.7 &  0.892 & 60.9 	 	& 531.7& 0.04821 & 53382  &  20.217\\
14 & 6 &  0.853 & 52.0 &  0.881 & 59.5 	 	& 531.9& 0.04801 & 53891  &  20.358\\
14 & 7 &  0.853 & 52.0 &  0.855 & 54.0   	 	& 526.8& 0.05039 & 48385  &  18.929\\
15 & 7 &  0.843 & 51.6 &  0.842 & 52.2 	 	& 528.1& 0.05071 & 47916  &  18.763\\
16 & 8 &  0.834 & 51.2 &  0.810 & 46.9 		& 524.2& 0.05297 & 43513  &  17.555\\
\hline
 \end{tabular}
\caption{Potential parameter $V^{\rm WS}_{0,s}, V^{\rm WS}_{0,p} $ and Pauli blocking shift 
$\Delta E_{A,Z}^{\rm Pauli}(P=0;T,n_B,Y_p) \approx  n_B\,\delta E_{A,Z} ^{\rm Pauli}(T)$, approximated by two interpolation fits.
 First version: $\delta E_{A,Z} ^{\rm Pauli}(T)  \approx A\, a_{AZ}\,\exp(-b_{AZ}T)$. 
Second version: $\delta E_{A,Z} ^{\rm Pauli}(T) \approx  A\, f_{AZ} /(T+g_{AZ} )^{3/2}$. Units: MeV, fm.}
\label{Tab:2}
\end{center}
\end{table}

For large numbers $A$, an appropriate description of the nucleon wave function is the shell model where 
the single-nucleon wave functions are obtained from an effective potential.
Within local potentials, a well-known example is the Woods - Saxon potential 
\begin{equation}
\label{WSwf}
 V^{\rm mf,WS}(r )= V^{\rm WS}_0/\left[1+e^{(r-R_A)/a}\right].
\end{equation}
Typical parameter values which reproduce the properties of heavy nuclei are  $V^{\rm WS}_0=52.06$ MeV, $R_A = 1.26 A^{1/3}$ fm, and $a = 0.662$ fm \cite{WoodsS}. 
However, the light nuclei are not well described by these fit parameter. 

To calculate Pauli blocking, we need the effective potential to reproduce the nucleon wave functions.
We take the general form (\ref{WSwf}) with parameters which reproduce the rms values. 
Within a variational approach, we consider the Gaussians as 
class of wave functions and find the parameter values of  (\ref{WSwf}) for which the solutions of the wave function reproduce 
the values of $B_s$ and $B_p=B_s\beta^{-1/2}$ presented in Tab. \ref{Tab:1}.
From the three parameter $V^{\rm WS}_0,R_A,a$ occurring in (\ref{WSwf}), we fix $R_A = 1.26 A^{1/3}$ fm and $a = 0.662$ fm as given above.
Thus, we consider $V^{\rm WS}_0$ as a fit parameter to reproduce the rms values of the corresponding nucleon wave functions. 
The solutions $V^{\rm WS}_{0,s},V^{\rm WS}_{0,p}$ are given for the $A \le 16$ nuclei in Tab. \ref{Tab:2}. 
The values $V^{\rm WS}_{0,s}$ are consistent with the value $V^{\rm WS}_0$ \cite{WoodsS} for the larger nuclei.
The values $B_s(A),B_p(A)$ are also shown in in Fig. \ref{fig:0}.
The decrease of  $B_s$ for increasing $A$ is given by our ansatz  $B_s(A) = 1.324\,A^{-1/6}$ fm$^{-1}$ and describes the smooth change of the $\alpha$-like 
$1s$ core with increasing $A$. The values of $B_p$ show strong deviations from $B_s$ for small $A< 10$.
This may be considered as a signature that the wave function of these exotic nuclei is not well described by the shell model as discussed in the subsequent section \ref{sec:clustering}.

Having the potential to our disposal, we can calculate the Pauli blocking shift of the cluster as the sum over the shift (\ref{paulnu}) of the single-nucleon states.
If we approximate the Fermi distribution by the classical distribution 
\begin{equation}
 f_\tau(1) \approx \frac{n_\tau}{2}\left( \frac{2 \pi \hbar^2}{m T}\right)^{3/2} e^{-\hbar^2 p_1^2/(2 m T)}
\end{equation}
valid in the low-density region ($\mu_\tau < 0$), a linear dependence on the baryon density results. In general we have
\begin{equation}
\label{delEP}
 \Delta E^{\rm Pauli}_{A,Z}(P;T,\mu_n,\mu_p)= \sum_\nu  \Delta E^{\rm Pauli}_{1,\nu}= n_B F_{A,Z}(Y_p)\, \delta E^{\rm Pauli}_{A,Z}(P;T) +{\cal O}(n_B^2).
\end{equation}
 For symmetric matter ($n_n=n_p$) follows $F_{A,Z}(Y_p)=1$, for asymmetric matter we have 
\begin{equation}
\label{Yp}
n_B\, F_{A,Z}(Y_p)= \frac{2}{A}(N n_n+ Z n_p). 
\end{equation}

We present here results for  $P=0$. For the light $1s$ elements, the $\bf P$ dependence is discussed in Ref.  \cite{r3}. 
According to Eqs. (\ref{paulnu}) and  (\ref{delEP}) we have 
$ \delta E^{\rm Pauli}_{A,Z}(T)=4\, \delta E^{\rm Pauli}_{AZ,s}(T)+(A-4)\, \delta E^{\rm Pauli}_{AZ,p}(T)$. 
The separate contributions of the 4 nucleons in the $s$ orbit
and the ($A-4$) nucleons in the $p$ orbit are given in the Appendix \ref{Mott}, Tab.~\ref{Tab:3}, for different $T$, see also Fig. \ref{fig:1}.
Interpolations for $\delta E^{\rm Pauli}_{A,Z}(T)$ are shown in Tab.~\ref{Tab:2}, see Sec.~\ref{Inter} below.  

A consequence of the Pauli blocking is that the in-medium binding energy of the cluster is decreasing. 
In \cite{RMS82} the Mott density $n^{\rm Mott}_{A,Z}(T)=B_{A,Z}/ \delta E^{\rm Pauli}_{A,Z}(P=0,T)$ has been introduced characterizing the density where the bound state is dissolved.
It depends on $T$ as shown in Tab.~\ref{Tab:3} of Appendix \ref{Mott}. In general, 
the Mott condition $B_{A,Z}-\Delta E^{\rm Pauli}_{A,Z}(P;T,\mu_n,\mu_p)=0$ gives a critical baryon density which depends not only on $T$ but also on $\bf P$ and asymmetry.
In the case of $^6$Li, we have the situation where the bound state dissolves into a (medium-modified) $\alpha$ particle and two nucleons. 
This leads to further reduction of the Mott density.

Note that the values for the Mott density given in Tab.\ref{Tab:3} cannot be interpreted such that any $A$-nucleon correlations disappear for increasing density at this value. Above the Mott density, bound states may exist for ${\bf P} \neq 0$ where the blocking is smaller, see the decreasing overlap with increasing $|\bf P|$ in Fig. \ref{fig:P}. In addition, correlations are present in the continuum, see Section \ref{Sec:5He} below, and contribute to the composition of nuclear matter above the Mott density \cite{SRS}.

\subsection{Cluster model}
\label{sec:clustering}

\begin{table}
\begin{center}
\hspace{0.5cm}
 \begin{tabular}{|c|c|c|c|c|c|c|c|}
\hline
$A$  &  $Z$  &$ \delta E^{\rm Pauli}_{A,Z}(5)/A$&$ \delta E^{\rm Pauli}_{A,Z}(20)/A$& $a_{AZ}$&$b_{AZ}$&$f_{AZ}$  &$g_{AZ}$\\
{} & {} & [MeV fm$^{3}$] & [MeV fm$^{3}$]&[MeV fm$^3$]& [MeV$^{-1}$]& [MeV$^{5/2}$ fm$^3$]& [MeV]\\
\hline
2* & 1   &  384.4 & 79.0  & 695.6 & 0.12216 & 8715.8 & 3.011\\
3* & 1  &  524.8 & 160.8 & 791.1 & 0.08550 & 23175  & 7.493\\
3* & 2  &  528.5 & 146.3 & 831.5 & 0.09412 & 19482  & 6.0765\\
4* & 2  &  662.5 & 241.9 & 931.3 & 0.07117 & 41092  & 10.670\\
4  & 2  &  597.3 & 252.1 & 796.1 & 0.06002 & 50621  & 14.291\\
6* & 3  &  569.8 & 187.6 & 834.7 & 0.07997 & 28374  & 8.545\\
6c & 3  &  526.5 & 194.4 & 737.5 & 0.07096 & 32603  & 10.673\\
7* & 3  &  603.5 & 207.1 & 869.1 & 0.0762  & 32990  & 9.406\\
7c & 3  &  566.3 & 212.9 & 787.5 & 0.06908 & 37030  & 11.237\\
7* & 4  &  605.0 & 200.9 & 882.9 & 0.07897 & 30926  & 8.776\\
7c & 4  &  567.9 & 206.7 & 799.6 & 0.07178 & 34554  & 10.480\\
9* & 4 &  588.8 & 215.0 & 827.9 & 0.07117 & 36527  & 10.670\\
9c & 4  &  531.0 & 224.0 & 707.7 & 0.06002 & 44996  & 14.291\\
\hline
 \end{tabular}
\caption{
 Cluster states: $A^*$ shifts according \cite{r1}.  
$A$c: Adapted $\alpha$ shift from shell-model calculation.  
Pauli blocking shift 
$\Delta E_{A,Z}^{\rm Pauli}(P=0;T,n_B,Y_p) \approx  n_B\,\delta E_{A,Z} ^{\rm Pauli}(T)$, approximated by two interpolation fits.
 First version: $\delta E_{A,Z} ^{\rm Pauli}(T)  \approx A\, a_{AZ}\,\exp(-b_{AZ}T)$. 
Second version: $\delta E_{A,Z} ^{\rm Pauli}(T) \approx  A\, f_{AZ} /(T+g_{AZ} )^{3/2}$. Units: MeV, fm.}
\label{Tab:2c}
\end{center}
\end{table}

The main issue to calculate the Pauli blocking is the knowledge of the many-nucleon wave function which determines the phase space occupation. 
The shell model is based on the concept of independent motion in a mean-field potential.
As a quasiparticle approach, correlations between the nucleons are neglected. 
However, this model is problematic for nuclei with small mass numbers. 

The  'clustered' elements Li, Be, B are weakly bound, $p$-shell nuclei which demand a special treatment. 
Whereas the ground states of C, N, O ($12 \le A \le 16$) may be reasonably approximated by a shell model, 
it fails for the lighter nuclei because of the strong clustering contribution to the ground state wave function. 
Clustering in nuclei is treated by the resonating group method (RGM) and related approaches, 
see \cite{Wild1977,Saito1977,Hori1977,Hori2012} and references given there.

A striking example is $^8$Be. In contrast to other light $n \alpha$ nuclei which are stable and have relative large binding energy,  
$^8$Be $(n=2)$ is unstable and decays in two $\alpha$ particles, see Tab. \ref{Tab:1}. The reason is the strong quartet clustering, and ab initio calculations \cite{Wiringa} show
a dumbbell-shaped intrinsic density distribution. Similar to the Hoyle state which also clearly shows a cluster structure, the THSR approach \cite{THSR} has been 
worked out to describe $\alpha$-like clustering in nuclei. 
Significant cluster structures are also observed in the neighboring nuclei $^7$Li, $^7$Be, $^9$Be using AMD (antisymmetrized molecular dynamics)  calculations \cite{Kanada}. 
For recent inelastic scattering see \cite{Egorov}.
Whereas $^9$Be can be discussed as a two-$\alpha$ bound state hold together by the additional neutron, 
$^7$Li and $^7$Be can be considered as bound state of $\alpha$ + $^3$H or $^3$He, respectively. 
Also $^6$Li may be contain deuteron-like correlations in addition to the $\alpha$ particle. 
The density distribution of the intrinsic ground state of $^9$Be is shown in Ref. \cite{Lyu}.

The wave function of the cluster model is given by the antisymmetrized product of the wave functions of the constituent subclusters. 
For instance, the THSR approach considers $^8$Be as antisymmetrized product of two $\alpha$-like Gaussians with two different width parameters 
describing the intrinsic motion and the center-of-mass motion of the constituent subclusters. 
The intrinsic density distribution of these exotic nuclei is characterized by two $\alpha$-like cluster for $^8$Be and $^9$Be. 
The Pauli blocking shift results mainly from the blocking of the intrinsic motion of these subclusters 
so that we approximate this by the sum of the Pauli blocking of the constituents.
Considering $^7$Be, $^7$Li and $^6$Li in the same way, we calculate the Pauli blocking shifts as the sum of the shifts of the constituent subclusters.
The shifts of the corresponding light clusters are given in \cite{r1} (denoted by asterisks). 
We used the expression (46) of Ref. \cite{r1} to calculate the Pauli blocking shifts of the constituents,
\begin{equation}
\label{fitPauli4}
  \Delta E^{\rm Pauli}_{A,Z}(P;n_B,Y_p,T)\approx n_B\, F_{A,Z}(Y_p) \, A\, \frac{f _{A,Z} }{(T+g_{A,Z})^{3/2}},
\end{equation}
with $F_{A,Z}(Y_p)  $ given by Eq. (\ref{Yp}). As before, we take $P=0$ neglecting the $\bf P$ dependence of the Pauli shift. 
Values for the parameter $f_{A,Z}$ and $g_{A,Z}$ are given in Tab. \ref{Tab:2c}.
We use the Gaussian approach for the wave functions, but take for consistency the shift of the $\alpha$ particle 
according to the present shell-model approach, Tab. \ref{Tab:2}, which slightly differs because the c.m. motion is not separated. 
These cluster values are denoted by $c$. They are used in our further discussion. Results are also shown in  Fig. \ref{fig:1}. 

In Appendix \ref{App:6Li} we check our cluster model approximation by considering the lightest exotic nucleus 
$^6$Li. A microscopic calculation is performed using an effective nucleon-nucleon interaction potential and separating of the c.m. motion. 
A large value of the Pauli blocking is obtained, see Tab. \ref{Tab:4}.
Note that the Pauli blocking is stronger for the cluster structure than the shell-model value. The wave function in the $1s$ state is large at $p=0$, but 
goes to zero for the $1p$ state so that the overlap with the Fermi distribution becomes small.

\begin{figure}[h]
\centerline{\includegraphics[width=350pt,angle=0]{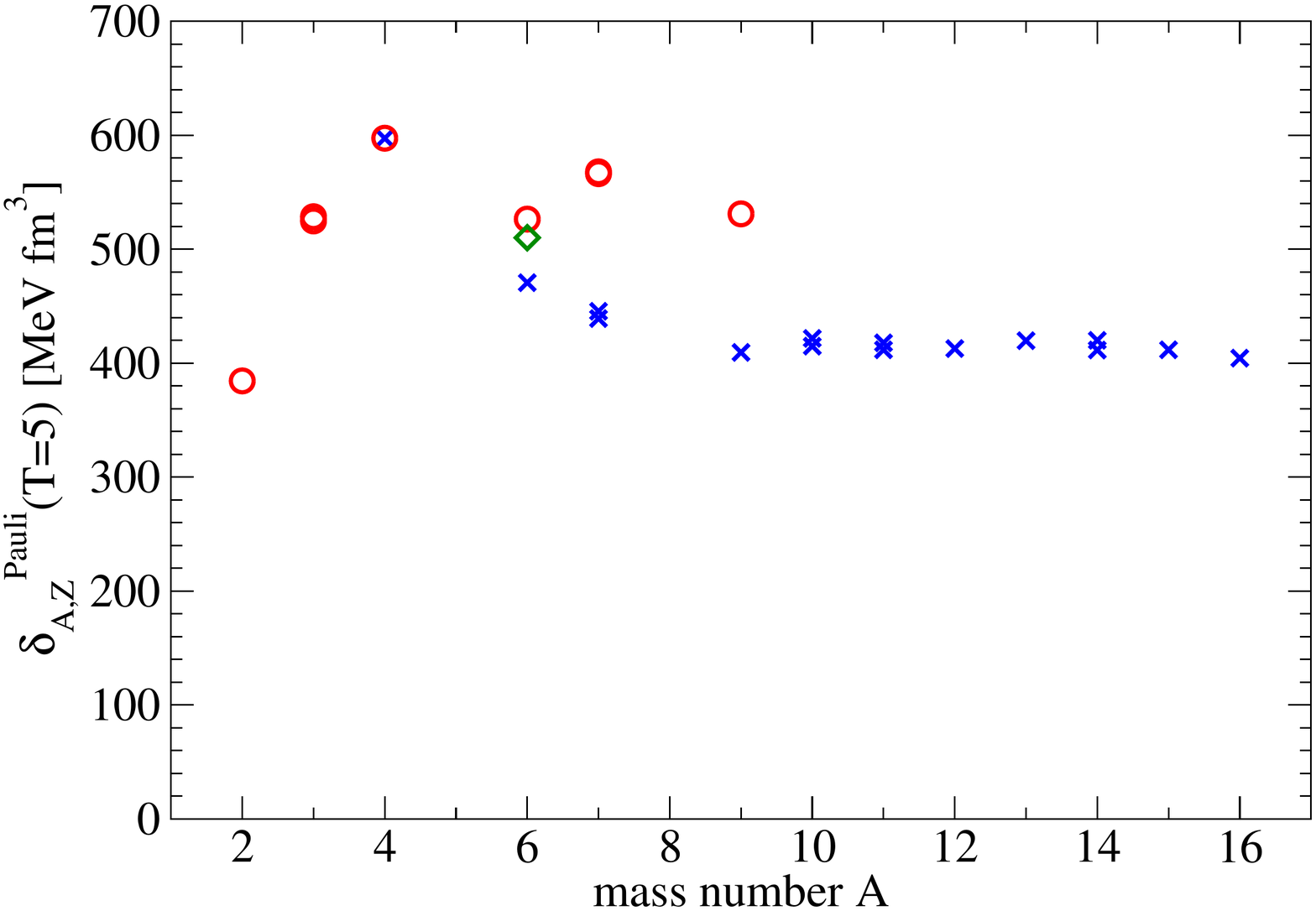}}
\caption{Pauli-blocking shift $\delta E_{AZ}^{\rm Pauli}(T)$ of $1s, 1p$ nuclei ($A \le 16$) at $T=5$ MeV. Shell model calculations (blue crosses) are compared to cluster model calculations (red circles). The green diamond gives the result of a microscopic calculation for $^6$Li. Values for $A < 4$ are taken from \cite{r1}.}
\label{fig:1}
\end{figure}

\subsection{Interpolation formula}
\label{Inter}
We consider the contribution to the energy shift, Eq.~(\ref{delEP}), which is linear in the baryon density. A calculation of the full density dependence of the energy shift has been performed for the deuteron~\cite{SRS} which shows that 
the contribution of correlations to the density is strongly suppressed above the Mott density. 
For the light elements, an expression for the contribution $\propto n_B^2$ has been given in Ref.~\cite{r3}.
As example, we calculate below in Eq.~(\ref{eeffHe}) the quadratic term $\propto n_B^2$ for the energy shift of $^5$He.
We suppose that the linear term of the energy shift is sufficient to describe the Mott effect. 
The higher order terms of the density expansion may become relevant near the saturation density.
They need a special treatment what is in general beyond the scope of the present work. 
We expect that near the saturation density any correlations beyond the quasiparticle approach are fading away.
A detailed description of this behavior is available at present only for some special cases such as $^2$H and $^5$He.

\begin{figure}[h]
\centerline{\includegraphics[width=350pt,angle=0]{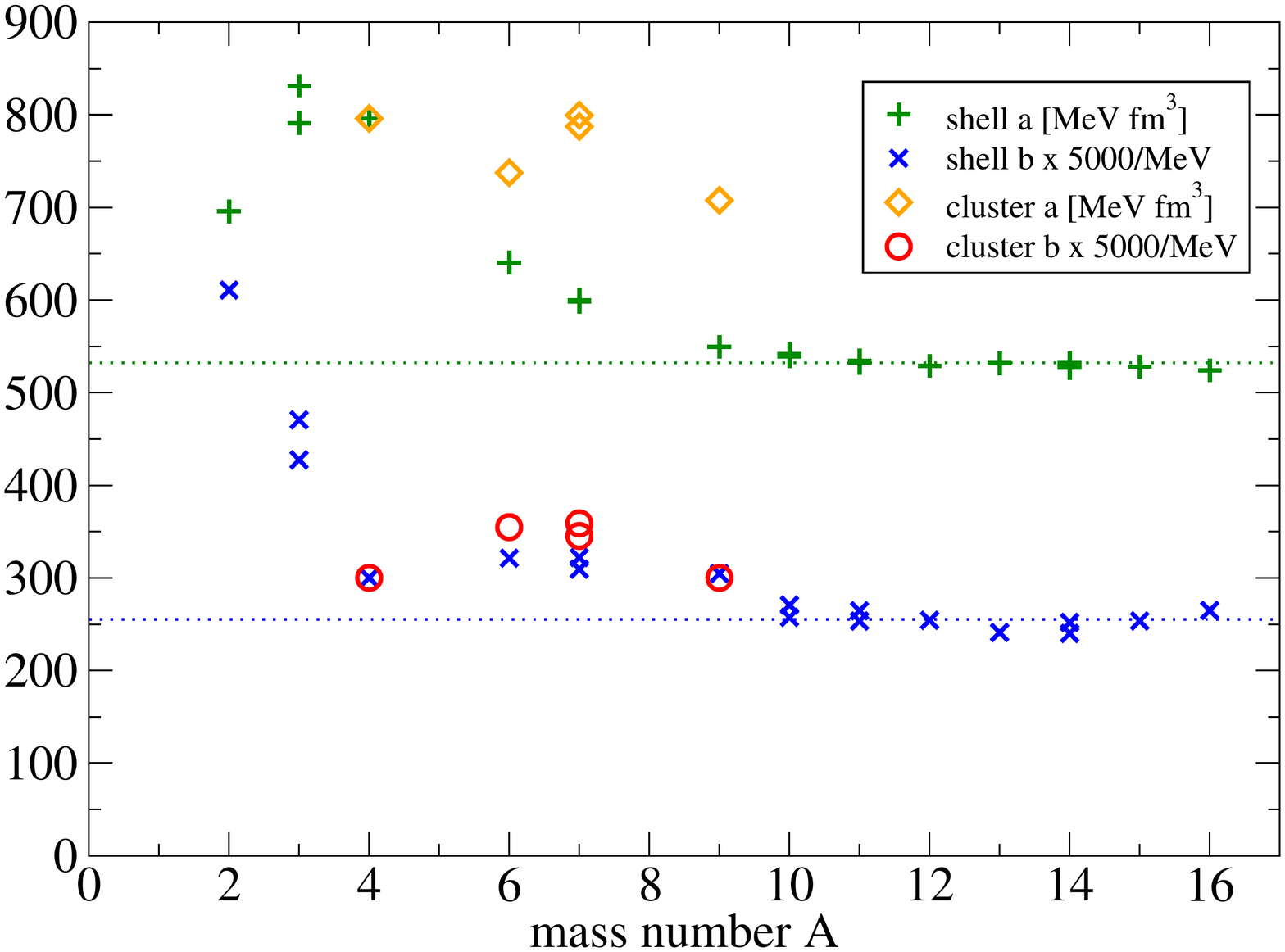}}
\caption{Parameter values of the fit (\ref{APauli}). The values $a_{A,Z}$ [MeV fm$^3$] and $5000 \times b_{A,Z}$ [MeV$^{-1}]$ are shown. 
Dotted lines are the averages $\bar a =532.0$ MeV fm$^3$ and $\bar b =0.05103$ MeV$^{-1}$.
Shell model calculations are denoted by green plus ($a_{A,Z}$) and blue crosses ($b_{A,Z}$), cluster model calculations by orange diamonds ($a_{A,Z}$) and red circles ($b_{A,Z}$).}
\label{fig:2}
\end{figure}

Calculations for the Pauli blocking shift $\delta E^{\rm Pauli}_{A,Z}(T)$ have been performed for all $p$-shell nuclei for 
$1\,{\rm MeV} \le T \le 20$ MeV and baryon densities up to the Mott density, see, e.g.,
Tab. \ref{Tab:3} in Appendix \ref{Mott}.
To implement the in-medium shifts in calculations of the composition of nuclear matter and related properties, 
we propose interpolation expressions for the Pauli blocking shift
\begin{equation}
\label{APauli}
  \Delta E^{\rm Pauli}_{A,Z}(P;T, n_B,Y_p)\approx n_B \, F_{A,Z}(Y_p) \, A\, \,a_{A,Z} \,\, e^{-b_{A,Z} T}
\end{equation}
where we neglect the dependence on $\bf P$ so taking $P=0$. 
With the results shown in Tab. \ref{Tab:3}, we obtain from a least square deviation fit the values $a_{A,Z}$ and $b_{A,Z}$ 
given in Tabs. \ref{Tab:2}, \ref{Tab:2c}, see also Fig. \ref{fig:2}. 
The relative deviations of the interpolation fit (\ref{APauli}) are below 2 \% in the region considered here.

A similar fit (\ref{fitPauli4}) has been proposed in \cite{r1} for the light cluster $A\le4$.
 We give also the values $f_{A,Z}$ and $g_{A,Z}$ in Tabs. \ref{Tab:2}, \ref{Tab:2c}. The relative deviations are below 4 \%.
 Within the parameter region discussed here, both interpolation formulas give similar results.
 However, outside this region, (\ref{fitPauli4}) overestimates the behavior at low temperatures where the phase space near ${\bf p}=0$ is relevant. 
 There, the $1p$ wave function has zero density so that Pauli blocking is less efficient. This lower value for the Pauli blocking is better reproduced by expression (\ref{APauli}).

Within this work, we use the fit (\ref{APauli}). The Pauli blocking shifts $ \Delta E^{\rm Pauli}_{A,Z}$ are nearly proportional to the mass number $A$.
The corresponding parameter values $a_{A,Z}$ and $5000 \times b_{A,Z}$ are shown in Fig. \ref{fig:2}.
For the nuclei $10 \le A \le 16$ we have the average values $\bar a= 532.0$ MeV fm$^3$ and $\bar b= 0.05103  $ MeV$^{-1}$. 
These values are also shown in Fig. \ref{fig:2} (dotted lines).
In conclusion, the expression
\begin{equation}
 \Delta E^{\rm Pauli}_{A,Z}(P;T,\mu_n,\mu_p) \approx  A \times 532.0\, e^{-0.05103\, T/{[\rm MeV]}} n_B [{\rm MeV\,fm}^3]
\end{equation}
works for 1$p$ nuclei with $10 \le A \le 16$. 

For the lighter, clustered nuclei the shell model approach considering a wave function formed by $1s, 1p$ orbitals is not applicable. 
The strong deviation of $a_{A,Z},\,b_{A,Z}$ from the average values $\bar a,\,\bar b$, respectively, are caused by the anomalous large rms radii of the rare element nuclei
and clustering effects.

\section{Equation of state including scattering states; loosely bound objects}
\label{Sec:virial}

\subsection{Generalized Beth-Uhlenbeck approach}

The equation of state (\ref{eos}) includes the sum over excited states $\nu$, in particular
the continuum of scattering states. Considering $1p$ nuclei, excited states are of relevance. As example, the contribution of the channel describing few-nucleon correlations
with  $A=8$ to the 
equation of state contains also the nucleus  $^8$Be. Sometimes it is included in NSE as a real nucleus, 
decaying quickly into two $\alpha$ particles. 

A more systematic quantum statistical approach
is necessary to treat continuum correlations. In the $S$-wave $\alpha+\alpha$ channel, $^8$Be appears as a resonance and contributes to the virial coefficient 
$b_\alpha$ investigated in Ref. \cite{HS}. Similarly, unstable nuclei in other channels should be treated as continuum correlations via the scattering phase shifts.
 The need to treat scattering states is evident when 
considering in-medium effects and the dissolution of bound states because of Pauli blocking.
The bound state contribution to the partial densities (\ref{eos}) shows a discontinuous behavior if a bound state merge with the continuum and disappears. 
This discontinuity is compensated taking into account continuum contributions according to the Levinson theorem.

To describe these effects, the intrinsic partition function $z^{\rm part}_{A,Z,J}({\bf P};T,\mu_n,\mu_p)$ 
of the channel $\{A,Z,J\}$ at ${\bf P}$ is introduced in \cite{r3},
\begin{eqnarray}
\label{components}
 n^{\rm part}_{A,Z,J}(T,\mu_n,\mu_p)= \int \frac{d^3 P}{(2 \pi)^3}e^{- \hbar^2 P^2/(2AmT)} e^{\left(N\mu_n+Z\mu_p\right)/T} z^{\rm part}_{A,Z,J}({\bf P};T,\mu_n,\mu_p)\,.
\end{eqnarray}

A further subdivision into a bound part and a continuum part, 
$z^{\rm part}_{A,Z,J}({\bf P};T,\mu_n,\mu_p)=z^{\rm bound}_{A,Z,J}({\bf P};T,\mu_n,\mu_p)+z^{\rm cont}_{A,Z,J}({\bf P};T,\mu_n,\mu_p)$,
is not free of ambiguity.
We choose as bound state contribution
\begin{eqnarray}
\label{zpart}
 z^{\rm bound}_{A,Z,J}({\bf P};T,\mu_n,\mu_p)&=&(2 J+1) e^{-E_{A,Z,J}^{\rm cont}({\bf P})/T}
\sum_{\nu}^{\rm bound} \,\,\left[e^{B_{A,Z,J,\nu}({\bf P};T,\mu_n,\mu_p)/T}-1\right]
 \,\,\Theta\left[B_{A,Z,J,\nu}({\bf P};T,\mu_n,\mu_p)\right]
\end{eqnarray}
where the in-medium binding energy is given by $B_{A,Z,J,\nu}({\bf P})=-E_{A,Z,J,\nu}({\bf P})+E_{A,Z,J}^{\rm cont}({\bf P})$. Here, $E_{A,Z,J}^{\rm cont}({\bf P})$ is the edge of continuum 
in the channel under consideration.
%$\Theta(x)=1$ for $x>0$ and zero for $x < 0$ denotes the step function.
We use already the quasiparticle approach where the density effects are taken into account in the mean-field approximation.
In particular, the  single-nucleon states are shifted by the self-energy. As example, we  use the parametrization \cite{TRB,r3}   of the relativistic mean-field approximation DD2-RMF \cite{Typel}. 
For the bound state energies $E_{A,Z,J,\nu}({\bf P};T,\mu_n,\mu_p)$, we take the solution 
of the in-medium Schr{\"o}dinger equation (\ref{Awave}) containing the single-particle shifts and the Pauli blocking (\ref{APauli}), see Ref. \cite{r1}.
The "-1" in the bound-state contribution (\ref{zpart}) is a relict of the scattering state contribution according to the Levinson theorem, see Eq. (\ref{intrd}) below. It makes the bound-state contribution 
continuous if the binding energy goes to zero. 
Here, the continuum edge of the cluster constituents at the same total momentum ${\bf P}$ is for the decay into single nucleons
\begin{eqnarray}
\label{Econt}
E^{\rm cont}_{A,Z,J}({\bf P};T,\mu_n,\mu_p)
&=&N E_n({\bf P}/A;T,\mu_n,\mu_p)+Z E_p({\bf P}/A;T,\mu_n,\mu_p).
\end{eqnarray}
A similar relations gives the edge of the continuum if other decay channels containing subclusters are considered.
The argument of the step function $\Theta(x)=1, x \geq 0;\,\,=0$ else, denotes the binding energy which must be 
positive to have a bound state. Above the Mott density, this condition is a restriction for the summation over 
 $\bf P$ to that region where bound states may exist. If the quasiparticle shift is taken in effective mass approximation,
 the shift can be transferred to the chemical potential. 

The contribution of two interacting clusters to the EoS is related to the scattering phase shifts according to 
Beth and Uhlenbeck \cite{BU,Huang}. For instance, for the deuteron channel $^2{\rm H}=d$ ($A=2, Z=1, J=1$) 
we have the generalized Beth-Uhlenbeck formula \cite{SRS}
\begin{equation}
\label{intrd}
z_{d}^{\rm part}({\bf P};T,\mu_n,\mu_p)= e^{- \frac{\hbar^2 P^2}{4mT}-\frac{E^{\rm cont}_d({P})}{T}}\,3\left[\left(e^{B_d(P)/T}-1\right) \Theta[B_d( P)]+ \frac{1}{\pi T}\int_0^\infty dE\ e^{-E/T} \left\{\delta_d(E)-\frac{1}{2} \sin [2 \delta_d(E)]\right\}\right]
\end{equation}
with medium-modified  bound state energies and phase shifts also obtained from the in-medium Schr{\"o}dinger equation (\ref{Awave}) to be consistent.
The generalized Beth-Uhlenbeck formula \cite{SRS} considers already the quasiparticle distribution so that the single-particle energies are shifted by 
a mean-field contribution.
Note that the term $-\frac{1}{2} \sin [2 \delta_d(E)]$ in Eq. (\ref{intrd}) compensates the contributions already used 
for the mean-field shift of the single-nucleon quasiparticle energies \cite{SRS}.
Thus, double counting of interaction terms is avoided.

In the low-density limit, the in-medium modifications can be neglected, and the Fermi/Bose distributions are replaced by the Boltzmann distribution. 
 In this ordinary Beth-Uhlenbeck formula for the second virial coefficient, the single-particle contribution is described by the distribution of free, non-interacting nucleons.
 The integral over ${\bf P}$ in Eq. (\ref{intrd}) can be performed and  
 \begin{equation}
\label{ndfree}
 n^{\rm part,0}_d(T)= 3 \left(\frac{2 m T}{2 \pi\hbar^2}\right)^{3/2} e^{(\mu_n+\mu_p)/T} 
\left[e^{-E^{(0)}_d/T}-1+ \frac{1}{\pi T}\int_0^\infty dE\ e^{-E/T} \delta^{(0)}_d(E)\right]
\end{equation}
results. 

The scattering phase shift $\delta^{(0)}_d(E)$ as function of the kinetic energy $E$ of relative motion is mainly given by $\delta_{^3{\rm S}_1}(E)$, for a more detailed discussion of the low-density limit see \cite{HS}. There, the full contribution to the spin-triplet channel contains also the phase shifts $\delta_{^3{\rm D}_1}(E)$ etc. Within the virial expansion, we have 
\begin{equation} n^{\rm part,0}_d(T)=4/\Lambda^3 e^{(\mu_n+\mu_p)/T} b^{0}_d(T),
\end{equation}
where $\Lambda^2=2 \pi \hbar^2/(mT)$ and
\begin{equation}
\label{bd0}
 b^{0}_d(T)=\frac{3}{\sqrt{2}} \left[e^{-E^{(0)}_d/T}-1+ \frac{1}{2\pi T}\int_0^\infty dE_{\rm lab}\ e^{-E_{\rm lab}/2T} \delta^{(0)}_d(E_{\rm lab})\right],
\end{equation}
if the single nucleon contribution is given by the free nucleons, 
$n^{\rm part,0}_\tau(T)=2\Lambda^{-3} e^{\mu_\tau/T}$. Here, $E_{\rm lab}$ is the energy of the projectile hitting the 
resting target. Continuum contributions to the cluster-second virial coefficient from nucleon-nucleon, nucleon-$\alpha$, and $\alpha - \alpha$ scattering phase shifts have been given in \cite{HS}.

The inclusion of scattering phase shifts between two components of the cluster $\{ A,Z,J\}$ is seen from the square brackets in Eqs. (\ref{intrd}), (\ref{ndfree}), (\ref{bd0})
and suggests to define the intrinsic channel partition function
\begin{equation}
\label{C}
 C_{A,Z,J}(P )=\sum_\nu^{\rm bound} \left(e^{B_{A,Z,J,\nu}(P )/T}-1\right) \Theta(B_{A,Z,J,\nu}(P ))+ \frac{1}{\pi T} \int_0^\infty dE\,e^{-E/T} \left\{ \delta_{A,Z,J}(E,P)-
\frac{1}{2} \sin[2 \delta_{A,Z,J}(E,P)]\right\}
\end{equation}
where $E$ is the c.m. energy. The integral part in (\ref{C}) describing the continuum contribution was denoted in \cite{r3} as residual second virial coefficient. 
Binding energies and scattering phase shifts contain in-medium corrections so that they depend, in general, on ${\bf P}, T,\mu_n,\mu_p$.
Calculating this expression, the artificial subdivision in bound and continuum contribution becomes obsolete. A generalized phase shift may be introduced containing contributions of negative $E$,
where at each bound state energy a jump of $\pi$ happens, see Ref. \cite{r3}.

In-medium corrections are treated within the generalized Beth-Uhlenbeck approach \cite{SRS} for the nucleon-nucleon system.  In this work, the treatment of light clusters \cite{r1,r2,r3} is extended to the $1p$ nuclei. We have to determine the medium modifications of the scattering phase shifts solving Eq. (\ref{Awave}). 
This in-medium Schr{\"o}dinger equation
contains a potential, and, as usual, we choose the potential to reproduce the free scattering phase shifts. 
We will use a separable potential which leads to simpler expressions for the Pauli blocking, see Appendix \ref{Sec:separabel}.

\subsection{$^5$He, no in-medium shifts}
\label{Sec:5He}

For equilibrium nuclear matter with low proton fraction $Y_p$, neutron rich nuclei are dominant. In particular, triton $t$ ($^3$H) is more abundant than helion $h$ ($^3$He). Also, the  neutron rich nuclei $^4$H, $^5$He, $^6$He, etc., may become relevant \cite{Yudin}. However, they are not stable. We have to consider the channels, for which they appear as resonances in the continuum of scattering states. 

In this subsection we focus on $^5$He. 
It belongs to the channel with $A=5, Z=2, J=3/2$ which
contains the contribution of the unstable nucleus. The binding energy $B_{^5{\rm He}}=27.56$ MeV \cite{nuclei} is smaller than the binding energy of $^4$He so that
$\Delta B_{^5{\rm He},\alpha n} = B_{^5{\rm He}}-28.3\, {\rm MeV}=-0.7356$ MeV.  It  decays as 
$^5{\rm He} \to \alpha + n$, the half-life is $7 \times 10^{-22}$ s.

The partial density of the $^5$He channel is (we consider the virial coefficient $b_{\alpha n}(T) $ for the $\alpha - n$ system \cite{HS})
\begin{eqnarray}
\label{n5Heb}
n_{^5{\rm He}}&=&16 \left(\frac{ m T}{2 \pi \hbar^2}\right)^{3/2} b_{\alpha n}(T) \, e^{(-E_\alpha+3 \mu_n+2 \mu_p)/T}.
\end{eqnarray}
Within NSE, the partial density of this unstable nucleus would be (degeneracy $2J+1=4$)
\begin{eqnarray}
n^{\rm NSE}_{^5{\rm He}}&=&4 \left(\frac{5 m T}{2 \pi \hbar^2}\right)^{3/2} e^{(3 \mu_n+2 \mu_p+B_{^5{\rm He}})/T} \nonumber \\ &=&\frac{n_n}{2} n_\alpha 4 \left(\frac{5}{4} \frac{2 \pi \hbar^2}{ m T}\right)^{3/2}
e^{-0.7356\,{\rm MeV}/T} = 4 n_\alpha \left(\frac{5}{4}\right)^{3/2} e^{\mu_n/T} e^{\Delta B_{^5{\rm He},\alpha n}/T}.
\end{eqnarray}
This partial density contributes to the total neutron density with the factor 3 and to the total proton density with the factor 2. 
For a bound state with bound state energy $E_{^5{\rm He}}= -B_{^5{\rm He}}=-27.56$ MeV we have for relation (\ref{n5Heb}) 
\begin{equation}
\label{banNSE}
 b_{\alpha n}^{\rm NSE}(T)= \frac{5^{3/2}}{4} e^{(-E_{^5{\rm He}}+E_\alpha)/T}.
\end{equation}

However, instead of the unstable nucleus, we have to treat the continuum contributions, in particular the phase shifts.
It is an advantage of the Beth-Uhlenbeck formula that the second virial coefficient can be expressed in terms of 
properties which are directly observed, avoiding the introduction of a potential. As given in \cite{HS}, 
\begin{equation}
\label{HSban}
b^{\rm BU}_{\alpha n}(T)= \frac{5^{1/2}}{\pi T} \int_0^\infty dE_{\rm lab}\,e^{-4E_{\rm lab}/5T} \delta_{\alpha n}^{\rm tot}(E_{\rm lab}).
\end{equation}
The relative energy is $(4/5)\, E_{\rm lab}$, the later is the energy of the neutron, the $\alpha$ is fixed. 
Scattering phase shifts for the different $\alpha-n$ channels are given in Ref. \cite{Hoop} and parametrized  in \cite{Arndt}, in particular (units: MeV):
\begin{equation}
 \delta_{{\rm P}_{3/2}}(E_{\rm lab})= {\rm arccot}[(0.1281-0.1095\, E_{\rm lab}+0.006794\, E_{\rm lab}^2
-0.000113\, E_{\rm lab}^3)/(0.043733\, E_{\rm lab}^{3/2})]
\end{equation}
which gives the main contribution to  $\delta_{\alpha n}^{\rm tot}(E)=2 \delta_{{\rm S}_{1/2}}+2 \delta_{{\rm P}_{1/2}}+4 \delta_{{\rm P}_{3/2}}+\dots$.
Results for the virial coefficient (\ref{HSban}), calculated with  $\delta_{\alpha n}^{\rm tot}(E)  $  
as well as  the main contribution $ 4 \delta_{{\rm P}_{3/2}} $ at different values of $T$ are  shown in Tab. \ref{Tab:A}, see \cite{HS}. 
There exist also correlations
in the other channels ($ \delta_{{\rm S}_{1/2}},\delta_{{\rm P}_{1/2}} $) which partially compensate each other, higher angular momenta give almost no contribution to the density.

For comparison, calculations of phase shifts with a square well potential $V(r )=-V_0 \Theta(a-r)$; $V_0=55$ MeV, $a=2$ fm, 
as well as separable potential given in Appendix \ref{Sec:separabel}, Eq. (\ref{seppot}), with $\lambda =670$ MeV fm$^3$, $\gamma = 1.791$ fm$^{-1}$ 
are also given in Tab.~\ref{Tab:A}. Both potentials  are quite different but reproduce nearly 
the same phase shifts in the parameter region under consideration. The corresponding virial coefficients coincide in good approximation.

\begin{table}
\begin{center}
\hspace{0.5cm}
 \begin{tabular}{|c|c|c|c|c|c|}
\hline
$T$&  $b^{\rm NSE}_{\alpha n}$ &  $b_{\alpha n}$ \cite{HS}&  $b_{\alpha n}$  \cite{HS}&  
$b^{\rm BU}_{\alpha n}$, sq. w. &  $b^{\rm BU}_{\alpha n}$ s. p.\\ 
{[}MeV{]} & {} & full & P$_{3/2}$-wave & P$_{3/2}$-wave & P$_{3/2}$-wave \\
\hline
1   & 2.68 & 1.51 & 1.73 & 1.73  & 1.75 \\
2   & 3.87 & 2.26 & 2.48 & 2.48  & 2.49 \\
3   & 4.37 & 2.57 & 2.78 & 2.77  & 2.78 \\
4   & 4.65 & 2.73 & 2.91 & 2.88  & 2.90 \\
5   & 4.83 & 2.81 & 2.97 & 2.92  & 2.95 \\
6   & 4.95 & 2.86 & 2.99 & 2.92 & 2.96 \\
7   & 5.03 & 2.89 & 2.99 & 2.90 & 2.94 \\
8   & 5.10 & 2.92 & 2.98 & 2.87 & 2.91 \\
9   & 5.15 & 2.93 & 2.96 & 2.82 & 2.87 \\
10 & 5.19 & 2.95 & 2.93 & 2.77 & 2.82 \\
12 & 5.26 & 2.97 & -       &  2.65 & 2.71 \\
14 & 5.30 & 2.98 & -       & 2.53 & 2.59 \\
16 & 5.34 & 3.00 & -       & 2.41 & 2.48 \\
18 & 5.37 & 3.00 & -       & 2.30 & 2.37 \\
20  & 5.39 & 3.00& -       & 2.19 & 2.26 \\
\hline
 \end{tabular}
\caption{$N-\alpha$ virial coefficient $b_{\alpha n}$, Eq. (\ref{HSban}). The results of \cite{HS} using empirical phase shifts are compared to the NSE expression (\ref{banNSE}) and 
the Beth-Uhlenbeck calculations with phase shifts from two model potentials, a square well potential (sq. w.)  and a separable potential (s. p.), considering only the contribution of the 
P$_{3/2}$ channel.}
\label{Tab:A}
\end{center}
\end{table}

There is a significant contribution of the ${\rm P}_{3/2}$ channel which allows to introduce a nuclear  
state at negative binding energy $\Delta B^{\rm eff}_{^5{\rm He},\alpha n}(T)$.
However, the NSE value $\Delta B_{^5{\rm He},\alpha n} =-0.736$ MeV overestimates the contribution of the $^5$He channel. 
In particular at high temperatures, the virial form gives lower values for the partial density what is also known from the deuteron case. 
There, the introduction of an effective energy to account for the contribution of the continuum was also proposed in Ref. \cite{VT}.
As a first comment if comparing our generalized virial approach to the NSE, 
the contribution of the $^5$He channel is essentially reduced, in particular at higher $T$.

\begin{figure}[h]
(a)\includegraphics[width=220pt,angle=0]{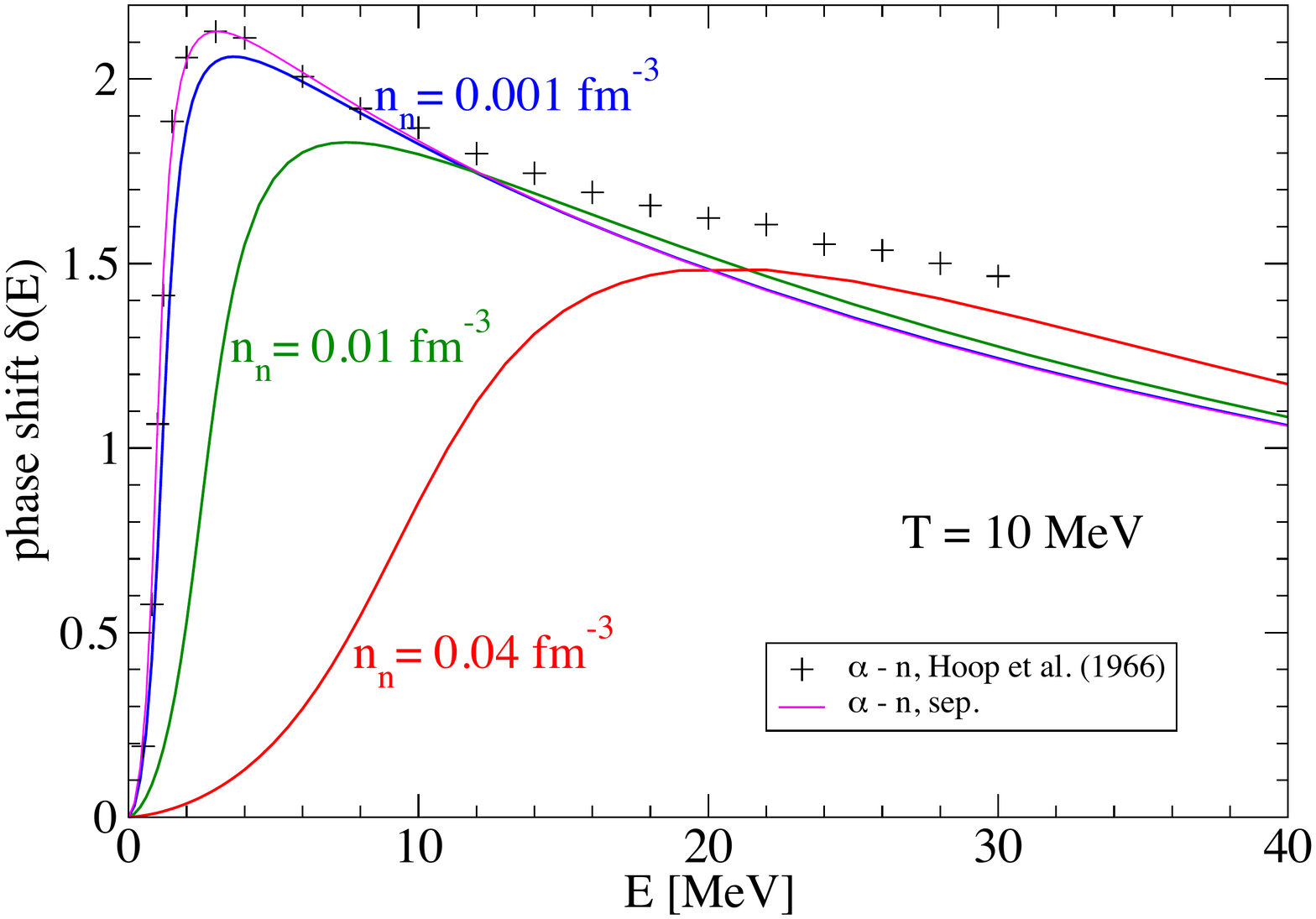}
(b)\includegraphics[width=220pt,angle=0]{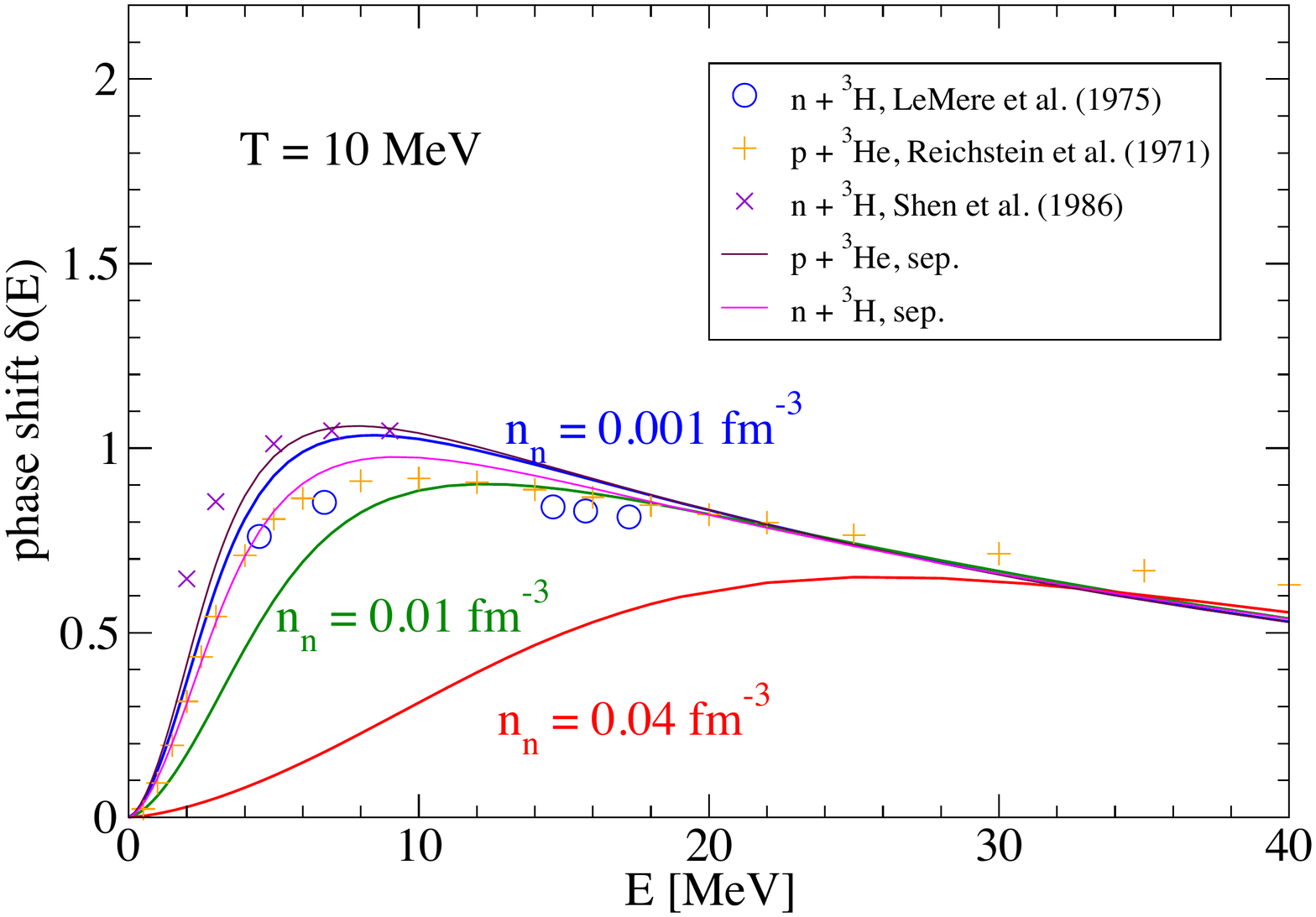}
\caption{In-medium scattering phase shifts for $\alpha - n$ (a) and $n - ^3$H, $p - ^3$He (b).
Experimental data of Hoop et al. \cite{Hoop}, Reichstein et al. \cite{Reichstein} and calculations of LeMere et al. \cite{LeMere}, Shen et al. \cite{Shen}  are compared to the separable potentials with parameters given in the text. Medium modifications are shown for $T= 10$ MeV and different free neutron densities $n_n$.}
\label{fig:phase}
\end{figure}

\subsection{$^5$He, with in-medium shifts}

As shown in Eq. (\ref{eos}), a cluster decomposition of the single-particle self-energy allows the decomposition of the total baryon density into partial densities. 
For $A>1$ we obtain the contribution of the channel ${\cal C}=\{A,Z,J\}$
\begin{equation}
\label{npartC}
 n^{\rm part}_{A,Z,J}(T,\mu_n,\mu_p)=(2J+1) \left(\frac{Am T}{2 \pi \hbar^2}\right)^{3/2} 
e^{\left(-E^{\rm cont}_{A,Z,J}+N\mu_n+Z\mu_p\right)/T} \,C_{A,Z,J}(T,\mu_n,\mu_p),
\end{equation}
neglecting quantum degeneracy and the $\bf P$-dependence of $C_{A,Z,J}({\bf P};T,\mu_n,\mu_p)$ (\ref{C})
so that the integral over the c.m. momentum $\bf P$ can be performed. Within a quasiparticle approach, 
in-medium bound-state energies and scattering phase shifts are used to evaluate the intrinsic channel partition function $C_{A,Z,J}$.
Note that in general the $\bf P$ dependence may be taken into account within an effective mass approximation.
The continuum edge $E^{\rm cont}_{A,Z,J}({\bf P};T,\mu_n,\mu_p)$  (\ref{Econt}) is also taken for $P=0$.
A binary effective interaction of subclusters ${\cal C}_1, {\cal C}_2$, with $ {\cal C} \rightleftharpoons {\cal C}_1+ {\cal C}_2$, is considered which leads to the scattering phase shifts 
$\delta_{{\cal C};{\cal C}_1{\cal C}_2}(E)$, where  $E$ denotes the intrinsic energy of relative motion of the subclusters. The continuum edge
$E^{\rm cont}_{{\cal C};{\cal C}_1,{\cal C}_2}(T,\mu_n,\mu_p)=E^{\rm qu}_{{\cal C}_1}(T,\mu_n,\mu_p)+E^{\rm qu}_{{\cal C}_2}(T,\mu_n,\mu_p)$ is obtained in the 
rigid shift approximation. We neglect the in-medium modification of the effective masses.
As discussed above, the binding energy in the special binary channel 
$\Delta B_{{\cal C};{\cal C}_1{\cal C}_2}(T,\mu_n,\mu_p)=B_{\cal C}-E^{\rm cont}_{{\cal C};{\cal C}_1{\cal C}_2}$
relative to the corresponding continuum edge of subclusters has to be taken. 
The single-nucleon contribution ($A=1$) follows from the quasiparticle shift calculated, e.g., from the DD2-RMF approach \cite{TRB}.
In general, the fermionic distribution function is used to calculate the single-nucleon densities.

We apply Eq. (\ref{npartC}) to the binary reaction channel $\alpha+n \rightleftharpoons\, ^5$He, $A=5, Z=2,J=3/2$.
We have
\begin{equation}
\label{npart5He}
 n^{\rm part}_{^5{\rm He}}(T,\mu_n,\mu_p)=4 \left(\frac{5}{4}\right)^{3/2} n_\alpha(T,\mu_n,\mu_p)e^{-\Delta E_n^{\rm SE}/T+\mu_n/T}
 C_{^5{\rm He};\alpha n}(T,\mu_n,\mu_p),
\end{equation}
with (there are no bound states)
\begin{equation}
\label{C5He}
 C_{^5{\rm He};\alpha n}(T,\mu_n,\mu_p)=\frac{1}{\pi T} \int_0^\infty dE\,e^{-E/T} \left\{ \delta_{^5{\rm He};\alpha n}(E)-
\frac{1}{2} \sin[2 \delta_{^5{\rm He};\alpha n}(E)]\right\}=\exp\left[-E^{\rm eff}_{^5{\rm He};\alpha n}(T,\mu_n,\mu_p)/T\right]
\end{equation}
with $E$ being the energy of relative motion. For the sake of parametrization, we introduce the effective channel energy
$E^{\rm eff}_{^5{\rm He};\alpha n}(T,\mu_n,\mu_p)$ which may be considered as an effective, medium-dependent excitation energy to describe the statistical weight of the corresponding channel. The in-medium scattering phase shifts contain the Pauli blocking effects, single-nucleon self-energies cancel with continuum contributions in the rigid shift approximation.

Compared to (\ref{HSban}), a more general Beth-Uhlenbeck (BU) result is \cite{SRS}
\begin{equation}
\label{genBU}
b^{\rm gBU}_{\alpha n}(T,\mu_n,\mu_p)= 4 \frac{5^{1/2}}{\pi T} \int_0^\infty dE_{\rm lab}\,e^{-4E_{\rm lab}/5T} \left\{ \delta_{\alpha n}(E_{\rm lab})-
\frac{1}{2} \sin[2 \delta_{\alpha n}(E_{\rm lab})]\right\}
\end{equation}
with in-medium phase shifts. The free neutron density is calculated from the Fermi distribution function containing the quasiparticle shift $\Delta E_n^{\rm SE}$ (as before, the $P$-dependence is neglected).

We calculate the values of $b_{\alpha n}(T,n_n)$ as function of the temperature $T$ 
and the free neutron density $n_n$ (only the motion of the neutron $1p$ orbit is blocked), using the BU formula, see (\ref{ndfree}), and the generalized BU expression  (\ref{genBU}), see (\ref{intrd}).
The partial density related to the channel $A=5, Z=2, J=3/2$ is 
\begin{equation}
\label{n5He}
 n_{^5{\rm He}}(T,\mu_n,\mu_p)=\frac{8}{\Lambda^3} b_{\alpha n}(T,n_n)\, e^{[-E_\alpha(T,\mu_n,\mu_p)+2 \mu_n+2 \mu_p]/T} \,e^{(-\Delta E_n^{\rm SE}+\mu_n)/T}.
\end{equation}
The free neutron density $n_n$ is obtained from the Fermi distribution function with given $T, \mu_n,\Delta E_n^{\rm SE}$. In-medium shifts for the $\alpha$ particle are taken from \cite{r1}, for the neutron shift $\Delta E_n^{\rm SE}$ we use the 
parametrization \cite{r3} of the DD2-RMF approximation \cite{Typel}. The calculations using a separable potential are given in Appendix \ref{Sec:separabel}.

We scan the region $1 \le T \le 20$ MeV and $n_n \le 0.1$ fm$^{-3}$. The results are  parametrized as follows (note that the account of continuum contributions by effective energies has also been considered in \cite{VT}):
\begin{equation}
\label{efhe}
  b_{\alpha n}(T,n_n) =\frac{5^{3/2}}{2} C_{^5{\rm He};\alpha n}(T,\mu_n,\mu_p)=\frac{5^{3/2}}{2} e^{-E_{\alpha n}^{\rm eff}(T,n_n)/T}.
\end{equation}
For practical use, the dependence of the effective energy on the baryon density is approximated as
\begin{equation}
\label{eeffHe}
E^{\rm eff}_{\alpha n}(T,n_n)= E_{\alpha n,0}(T)+E_{\alpha n,1}(T)\,n_n+E_{\alpha n,2}(T)\,n_n^2.
\end{equation}
Data presented in  Tab. \ref{Tab:P4} are reproduced with relative accuracy below 5 \% by the parameter values (units: MeV, fm)\,\,
\begin{eqnarray}
\label{Ean}
E_{\alpha n,0}(T)&=&0.85503+0.21729 \,T+0.031362 \,T^2; \nonumber \\
E_{\alpha n,1}(T)&=&100.05+80.749 \ln T+384.08 \exp(-0.44383 \,T); \nonumber \\
E_{\alpha n,2}(T)&=&-322.53+450.2 \ln T.
\end{eqnarray}

\subsection{$^4$H}
\label{Sec:4H}

Of interest is the $^4$H cluster in neutron-rich stellar matter.   It belongs to the channel $A=4,Z=1,J=2$. Similar to $^5$He, it is not bound, and appears as  correlations in the continuum of the $^3$H + $n$ channel. 
Measured phase shifts for $^3$H + $n$ are given in Ref. \cite{Seagrave}, see also \cite{LeMere,Shen}. Of interest are the $^3\delta_1(E)$ phases 
as function of the c.m. energies $E$ which are reproduced approximately by a separable potential (\ref{seppot}) with $\lambda = 1144.9$ MeV fm$^3$, $\gamma = 1.326$ fm$^{-1}$.
%Minimum, near 0, at 4.22 MeV, $S_n =-1.6$ MeV.

The mirror cluster $^4$Li appears in the $^3$He + $p$ channel and is more extensively studied, see \cite{Reichstein}. The corresponding virial coefficients have been considered in Ref. \cite{Connor},
see also \cite{Horowitz12}.
Here, the $^3\delta_1(E)$ phases 
as function of the c.m. energies $E$  are reproduced approximately by a separable potential with $\lambda = 967.9$ MeV fm$^3$, $\gamma = 1.377$ fm$^{-1}$. 
As in the case of the mirror nuclei $^3$H and $^3$He, the potential is weaker as in the  case $^3$H + $n$ because of Coulomb repulsion. %Nahe 0 bei 4.55 MeV, $S_p=-3.10$ MeV.

We calculate according Appendix \ref{Sec:separabel}
\begin{equation}
\label{C4HBU}
 C_{^4{\rm H};tn}(P=0,T,\mu_n,\mu_p)=\frac{1}{\pi T} \int_0^\infty dE e^{-E/T} \left\{ \delta_{^4{\rm H};tn}(E;P=0,T,\mu_n,\mu_p)-
\frac{1}{2} \sin[2 \delta_{^4{\rm H};tn}(E;P=0,T,\mu_n,\mu_p)]\right\}.
\end{equation}
For parametrization, the effective energy $ E_{tn}^{\rm eff}(P=0,T,\mu_n,\mu_p)= -T \ln[C_{^4{\rm H};tn}(P=0,T,\mu_n,\mu_p)]$  is introduced, see Tab. \ref{Tab:CH4}.
The data in the Table \ref{Tab:CH4} are reproduced with relative accuracy below 2\% by the approximation
\begin{equation}
\label{eeffH}
E^{\rm eff}_{t n}(T,n_n)= E_{t n,0}(T)+E_{t n,1}(T)n_n
\end{equation}
(higher order terms in $n_n$ can be neglected), the parameter values are (units: MeV, fm)
\begin{eqnarray}
\label{Etn}
E_{t n,0}(T)&=&3.0014+1.69165\, T+0.025471\, T^2; \nonumber \\
E_{t n,1}(T)&=&334.62+97.424 \ln T+356.09 \exp(-0.91636 \,T).
\end{eqnarray}

For the partial density of $^4$H we find
\begin{equation}
\label{npart4H0}
 n^{\rm part}_{^4{\rm H}}(T,\mu_n,\mu_p)=\frac{5}{2} \left(\frac{4}{3}\right)^{3/2} n_t(T,\mu_n,\mu_p)e^{-\Delta E_n^{\rm SE}/T+\mu_n/T}
 C_{^4{\rm H};tn}(T,\mu_n,\mu_p).
\end{equation}
There is another channel $A=4,Z=1,J=1$ of $^4$H containing the excited state $^4$H$^*$ at 0.31 MeV excitation energy. 
In our approach to consider the scattering phase shifts we have to consider the $^1\delta_1(E)$ phases and find the contribution
\begin{equation}
\label{npart4H}
 n^{\rm part}_{^4{\rm H}^*}(T,\mu_n,\mu_p)=\frac{3}{2} \left(\frac{4}{3}\right)^{3/2} n_t(T,\mu_n,\mu_p)e^{-\Delta E_n^{\rm SE}/T+\mu_n/T}
 C_{^4{\rm H}^*;tn}(T,\mu_n,\mu_p).
\end{equation}
with the corresponding phase shifts, describing the known values \cite{Reichstein}. Because the differences are small, 
we can assume that both contributions to the partial density can be put together. 
As before, we parametrize the result for the intrinsic channel partition function for both contributions as
\begin{equation}
\label{npart4Hexc}
 n^{\rm part}_{^4{\rm H}}(T,\mu_n,\mu_p)\approx 4 \left(\frac{4}{3}\right)^{3/2} n_t(T,\mu_n,\mu_p)e^{-\Delta E_n^{\rm SE}/T+\mu_n/T}
 C_{^4{\rm H};tn}(T,\mu_n,\mu_p),
\end{equation}

\begin{equation}
\label{C4H}
 C_{^4{\rm H};tn}(T,\mu_n,\mu_p)=
 \exp[E^{\rm eff}_{^4{\rm H};tn}(T,\mu_n,\mu_p)/T].
\end{equation}
There are also negative phase shifts $^1\delta_0(E),^3\delta_0(E)$ belonging to other channels, and affect the bound $1s$ nucleon, 
i.e. the medium modification of the triton. They lead to the negative values of the virial coefficient given in \cite{Connor} but will not be discussed here.

In contrast to $^5$He, the effective excitation energies for $^4$H are large and the values for the scattering phase shifts in the corresponding channels small. They are stronger influenced by the introduction of the quasiparticle picture [the sin term in Eq. (\ref{C4HBU})] 
and the Pauli blocking effects, as seen by the increase of the the effective excitation energies with density and temperature.

\section{Exemplary calculation of the composition of asymmetric nuclear matter}
\label{Sec:calculations}

\subsection{Virial, excited states and continuum correlations}

Having the partial densities $n^{\rm part}_{c}(T,\mu_n,\mu_p)$ of  the component $c$ to our disposal, 
the composition of nuclear matter is described by the mass fractions $X_c = A_c n^{\rm part}_c/n_B$  with $\sum_c X_c =1$.
This composition  is of interest for different applications such as HIC or astrophysical simulations.
For instance, the role of the lightest $p$ nuclei in the composition and the EoS has been discussed recently \cite{Yudin}. %The question behind is the role of correlations.
Such correlations determine the neutrino opacity, but the inclusion of the lightest $p$ nuclei to evaluate  
the  EoS and the composition demands special attention. 

We evaluate the composition of nuclear matter  for parameter values $T=10$ MeV, $Y_p=0.1$, as function of the mass density 
$\rho$ ($n_B=0.0597015 \times 10^{-14} \rho$ cm$^3$/g/fm$^3$, saturation baryon number density $n_{\rm sat}=0.15$ fm$^{-3}$, 
mass density $\rho_{\rm sat}=10^{14.4}$ g cm$^{-3}$).
 We consider the subsaturation density region ${\rm Log} [\rho]=^{10}\!\!\log \rho [{\rm g/cm}^3] = 11 - 14$.
Asymmetric matter with a small value of $Y_p$ is of interest in stellar processes.
As shown in  \cite{Yudin}, Fig. 5, higher clusters (metals, $Z > 2$) are  not relevant in this region. 
Using standard approaches such as the NSE and the excluded volume model \cite{Hempel}, 
a large mass fraction of neutron-rich, unstable H and He isotopes is predicted there near the saturation density.

In order to discuss the account of in-medium corrections, we compare different approximations starting from a simple NSE approach. 
As result we show that the contribution of unstable, neutron rich isotopes to the composition is strongly reduced near the saturation density if in-medium effects are taken into account.

\begin{figure}[h]
\includegraphics[width=350pt,angle=0]{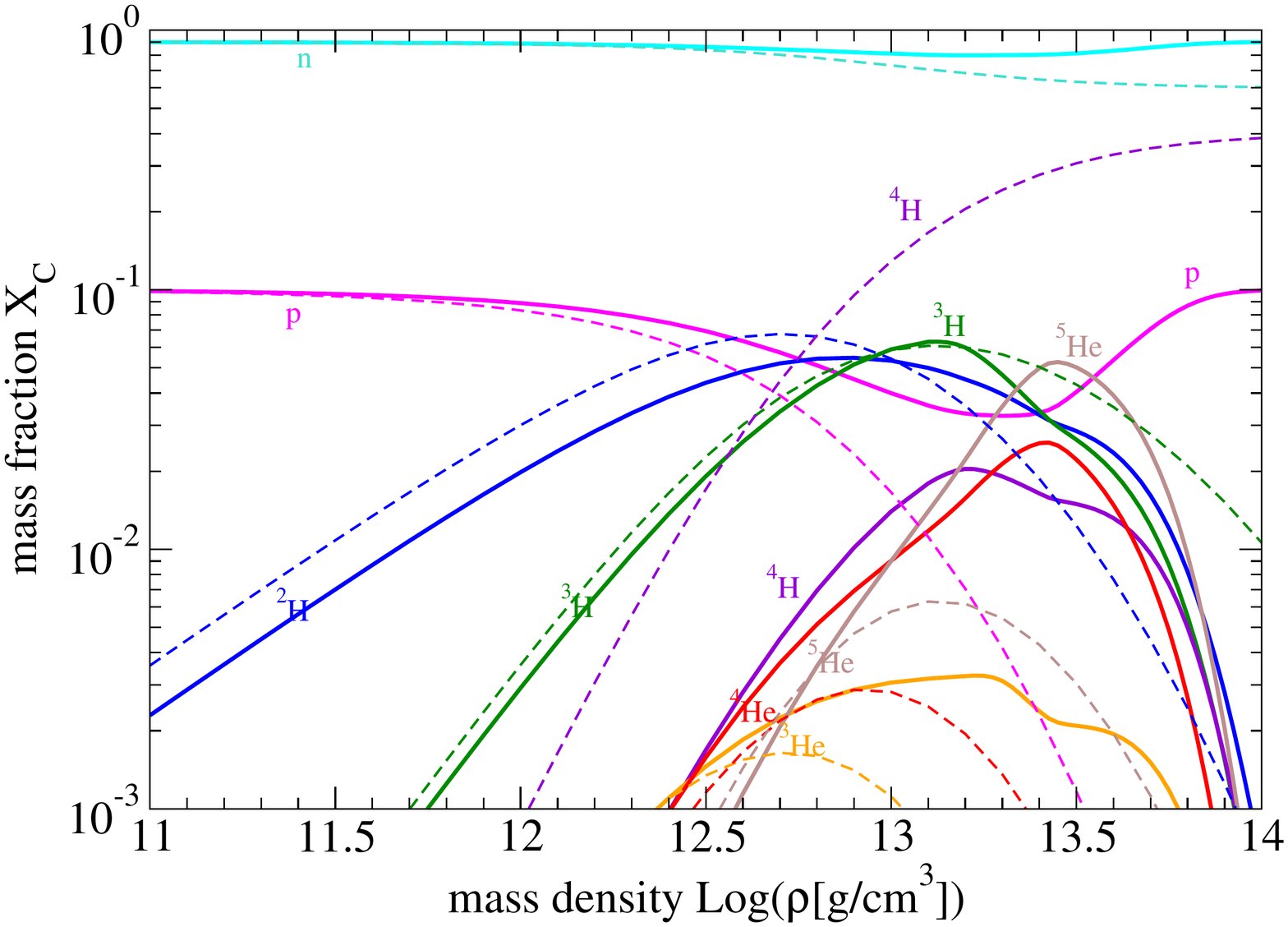}
\caption{Composition of nuclear matter, $T=10$ MeV, $Y_p=0.1$, as function of the mass density  including  $^4$H and $^5$He. 
The NSE without excited states (dashed) is compared to our quantum statistical in-medium approach (full lines).}
\label{fig:NSEQS6}
\end{figure}

The simple NSE model neglects all interaction beyond reactive collisions 
to establish the chemical equilibrium of the components. The large asymmetry $Y_p=0.1$ prefers the formation of neutron-rich clusters. 
Considering only the nearly stable elements $n,p,d,t,h,\alpha$, 
at high densities (Log$[\rho] > 13$), $t$ becomes dominant, almost all protons are bound to $t$. 
The inclusion of the subsequent neutron-rich, unbound isotopes $^4$H and $^5$He with known binding energies \cite{nuclei} into the NSE
leads to changes in the high-density region, see Fig. \ref{fig:NSEQS6}, dashed lines.

Instead of $t$, $^4$H becomes dominant. In the density region Log$[\rho] > 13$ almost each proton is bound to this neutron-rich isotope. 
Similarly, $^5$He becomes larger than the $\alpha$ mass fraction. 
In the NSE calculation, known excited states of the nuclei according to the tables \cite{nuclei} are taken into account. 
In particular, there exists a low-lying level of $^4$H at 0.31 MeV excitation energy and $J=1$.

We can continue to include even further known isotopes such as $^5$H, $^6$H, $^7$H, $^6$He, $^7$He, $^8$He, etc., 
with known binding energies and degeneracy found in the tables \cite{nuclei}. As shown in \cite{Yudin}, Fig. 5, for the NSE model 
as well as for the excluded volume model, according to the mass action law the effect of the dominance of neutron rich isotopes becomes 
even more visible, and matter near saturation density Log$[\rho] \approx 14$ appears as a mixture of free neutrons, $^7$H,  $^8$He, and 
other unbound nuclei. 
It is evident that a model of non-interacting, unbound nuclei is not adequate to describe matter near the saturation density.
We will show in the subsequent sections that the mass fractions of all these weakly bound nuclei are strongly suppressed near the saturation density if in-medium corrections are taken into account.
 % \begin{figure}[h]
% \includegraphics[width=200pt,angle=0]{NSE10T01NSE.pdf}
% \includegraphics[width=200pt,angle=0]{NSE10T01NSE+n.pdf}
% \caption{Composition of nuclear matter, $T=10$ MeV, $Y_p=0.1$, as function of the mass density. 
% A simple NSE approach considering only the nearly stable elements $n,p,d,t,h,\alpha$ (left) is compared with a calculation taking into account also $^4$H and $^5$He (dashed) as well as their excited levels (full lines).}
% \label{fig:nse+n}
% \end{figure}
We can also add to the NSE further known nuclei with $Z > 2$. 
However, the mass fractions are very small (below $10^{-3}$) at the parameter values for $T, Y_p$ considered here. 

Surprisingly, also in the low-density limit the NSE model fails to describe correctly the composition of the system 
because, like excited states, also scattering states have to be included.
Unbound states are not stable and appear as a resonance in the continuum of scattering states. Above we referred to $^8$Be which appears in the two-$\alpha$ continuum. 
Instead of considering a resonance gas, we have to treat the continuum of scattering states consistently.

A systematic treatment not only of the contribution of excited states, but of the whole continuum of scattering states is given by the Beth-Uhlenbeck formula, see Eq. (\ref{ndfree}) for the case $A=2$.
In the channel $A=2,Z=1, J=0,1$	the contribution of the scattering states has been parametrized
over the temperature range $1 \le T \le 20$ MeV, see  \cite{Horowitz12}.
The second virial coefficient $ b_{pn}(T)$ contains 
the contribution of the deuteron bound state, but is reduced owing to the negative contributions of the continuum, 
in particular at increasing $T$. Another parametrization is given in \cite{r2}.
Virial coefficients for other channels such as $^4$Li, $^4$H are found in \cite{Connor}.

We introduce the virial terms for $d$,  $^4$H, and $^5$He
as described in Sec. \ref{Sec:virial}.
There is a significant reduction of $d$ and a large reduction of  $^4$H. The reason is the small binding energy so that 
most of the partial density  is determined by the integral over the continuum, in particular at increasing $T$.
Because the phase shifts for $t - n$ scattering are small, a strong reduction is obtained from the virial coefficient. 
A reduction is also observed for $^5$He, but the $\alpha - n$ phase shifts are rather large so that only a minor reduction results. 
For a more detailed discussion of the common treatment of bound state contribution and scattering states see also Refs. \cite{r2,r3}.

\subsection{Contribution of unbound nuclei $^4$H and $^5$He}
\label{unbound}

To investigate the contribution of the unbound nuclei $^4$H and $^5$He to the composition of nuclear matter ($T=10$ MeV, $Y_p=0.1$),
we use the expressions for the corresponding channels given in Sec. \ref{Sec:virial}, (\ref{npart5He}), (\ref{C5He}) for the $^5$He channel
and (\ref{npart4Hexc}), (\ref{C4H}) for the two $^4$H channels.
 As function of density, we show the composition of these quantum statistical calculations in Fig. \ref{fig:NSEQS6}.
An important feature of the account of in-medium effects, in particular Pauli blocking, is the suppression of the cluster mass fractions at near-saturation densities.
The mass fractions of $n,p$ increase to the values 0.9 and 0.1, respectively. 
This is already seen if the quantum statistical treatment of only the light $1s$ nuclei $d,t,h,\alpha$  is compared to the NSO approach  \cite{r1}. 
%This, including the reduction owing to the continuum contributions, has already been discussed earlier \cite{r1}.

Considering  the unbound nuclei $^4$H and $^5$He, the position of the edge of continuum should be 
correctly taken into account. With $C_{^4{\rm H}}^{(0)}(T)$ being the phase shift integral (\ref{C4H}) in the zero-density limit, omitting the  quasiparticle shift,
we have the virial form (we put $^4$H and $^4$H$^*$ together, see Eq. (\ref{npart4Hexc}))
\begin{equation}
n^{(0)}_{^4{\rm H}}= \frac{(5+3)\times 8}{\Lambda^3} e^{(-E_t+3 \mu_n+\mu_p)/T}C_{^4{\rm H}}^{(0)}(T))= \frac{32}{ 3^{3/2}}n^{(0)}_t e^{ \mu_n/T}C_{^4{\rm H}}^{(0)}(T),
\end{equation}

\begin{equation}
n^{(0)}_{^5{\rm He}}= \frac{4\times 5^{3/2}}{\Lambda^3} e^{(-E_\alpha+3 \mu_n+2 \mu_p)/T}C_{^5{\rm He}}^{(0)}(T)= 
\frac{5^{3/2}}{2}n^{(0)}_\alpha e^{ \mu_n/T}C_{^5{\rm He}}^{(0)}(T).
\end{equation}

Within the quantum statistical approach we calculate the mass fractions of the corresponding channels  according Eqs. (\ref{npart4Hexc}), (\ref{n5He}).
The in-medium effects are incorporated as quasiparticle shift of $t$ and $\alpha$ \cite{r3} so that the in-medium densities $n_t, n_\alpha$ appear, 
the neutron chemical potential is shifted, see Appendix \ref{App:SE}, and the in-medium expression $C(T)$, Eq. (\ref{C4H}), is taken.

The contribution of the unbound nuclei $^4$H and $^5$He is reduced at high densities so that the Mott effect becomes visible.
Near the saturation density the mass fraction of bound clusters are decreasing  so that $X_n, X_p$ approach 
the free quasiparticle limit 0.9, 0.1, respectively.
The reduction of weakly bound states originates from the 
contribution of scattering states as known from the deuteron channel. In addition, the introduction of the quasiparticle description
leads to a further reduction owing to the sin term (\ref{C}), since part of the scattering phase shift (Born approximation) is already 
taken into account by the self-energy shift of the single-nucleon states. Pauli blocking in dense matter reduces further
the contribution of the unbound nuclei. 
%Note that the scattering phase shifts are weaker than for the di-neutron state.
As consequence, we conclude that the NSE with additional account of these unstable nuclei largely overestimates their contribution
near the saturation density as seen in Fig. \ref{fig:NSEQS6}. 

\subsection{Leight $p$-shell nuclei in the EoS}

Medium effects have a significant influence on the abundances of exotic, light $p$-shell clusters in dense matter, in particular 
calculating the composition in thermodynamic equilibrium, Eq. (\ref{eos}). Within the quantum statistical approach, the partial densities of the 
leight $p$-shell nuclei are calculated with the in-medium energies 
\begin{equation}
E_{A,Z,J,\nu}({\bf P};T,\mu_n,\mu_p)= E^{(0)}_{A,Z,J}+\frac{\hbar^2P^2}{2 A m}+ 
\Delta E^{\rm Pauli}_{A,Z}({\bf P};T,\mu_n,\mu_p)+\Delta E^{\rm SE}_{A,Z}({\bf P};T,\mu_n,\mu_p).
\end{equation}
For the Pauli blocking term $ \Delta E^{\rm Pauli}_{A,Z}(P;T, n_B,Y_p)$ we use the interpolation formula (\ref{APauli}) as described in Sec. \ref{Inter}.
The self-energy term $\Delta E^{\rm SE}_{A,Z}({\bf P};T,\mu_n,\mu_p)$ is taken as the sum over the self-energy shifts $\Delta E^{\rm SE, RMF}_\tau(T=0,n_B,Y_p)$
of the constituting nucleons, taken here in the DD2-RMF approximation \cite{TRB,r3}, see also \cite{r3}. Because the self-energy shift 
is present in the bound states as well as the scattering states, it is of minor relevance for the Mott effect. As discussed in the Appendix \ref{App:SE},
a reduction is expected for heavy nuclei so that we take the self-energy shift with a factor 0.5. Up to the Mott density such additional effects are not essential.
Above the Mott density, a further reduction of the contribution of clustered, light $p$-shell clusters is expected because we have to treat, instead of bound states,
scattering states in the continuum as already presented in the previous Sec. \ref{unbound}. 

\begin{figure}[h]
\includegraphics[width=350pt,angle=0]{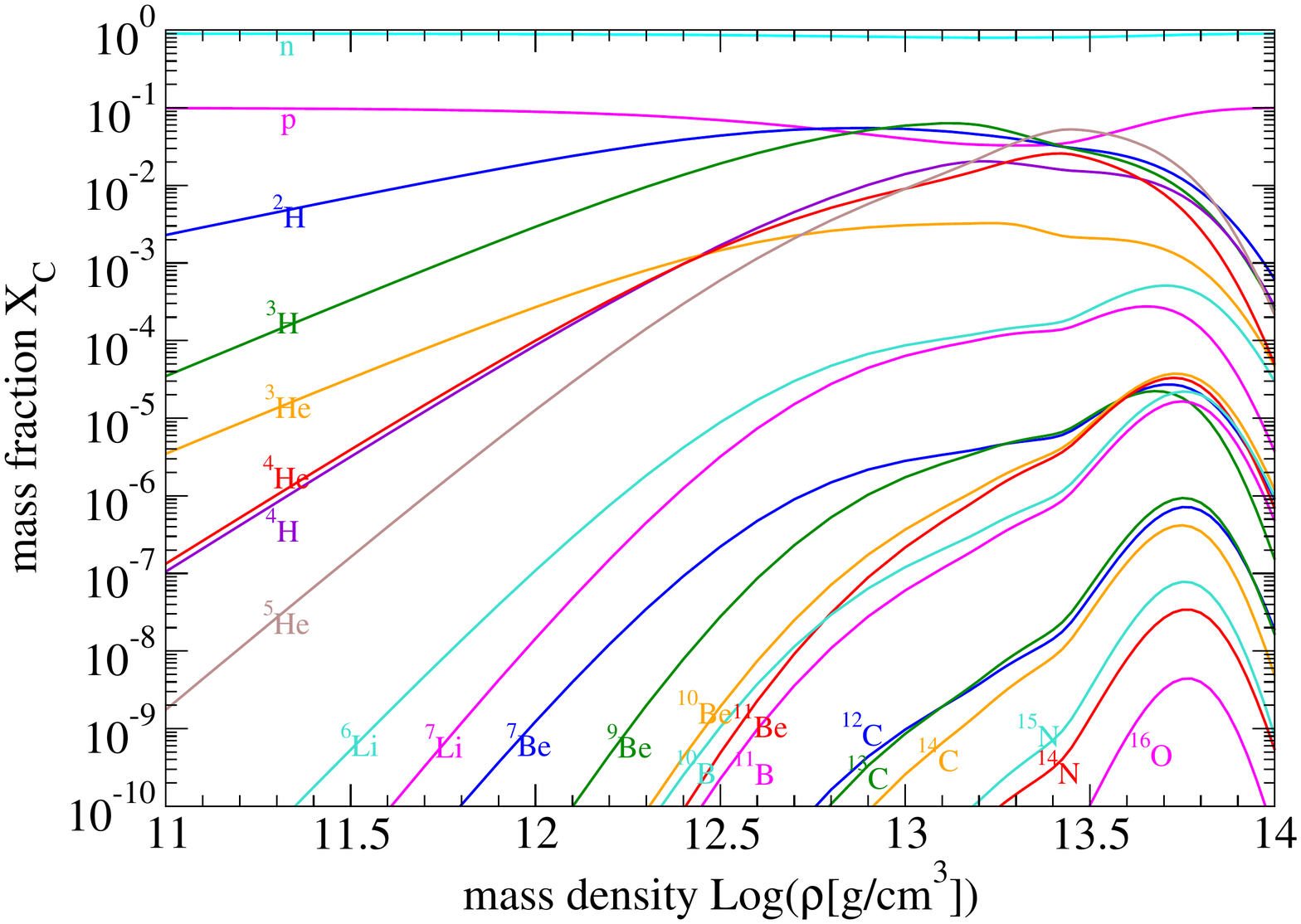}
\caption{Composition of nuclear matter, $T=10$ MeV, $Y_p=0.1$, as function of the mass density. 
Continuum edge with self-energy shift according to DD2-RMF at $P=0$, light elements ($A \le 4$) according \cite{r3}, 
Pauli blocking in scattering phase shifts by interpolation formula (\ref{eeffHe}),
(\ref{eeffH}), 
Pauli blocking of  light $p$-shell nuclei according (\ref{APauli}).}
\label{fig:QS16halbSE}
\end{figure}

As example, we show in Fig. \ref{fig:QS16halbSE} calculations for the conditions given above,
i.e. $T=10$ MeV and $Y_p = 0.1$ for mass densities $10^{11} \dots 10^{14}$ g/cm$^3$.  
Only the ground states of the nuclei with $A \le 16$ are considered. 
If comparing the quantum statistical approach to the ordinary NSE,
below $10^{12} $ g/cm$^3$ both 
approaches agree quite well, medium effects such as self-energy shifts and Pauli blocking are small. There are deviations 
because of the virial coefficient containing the continuum contributions, in particular for weakly bound nuclei such as $d$. 
The reduction of the abundance of deuterons at high temperatures has been discussed elsewhere \cite{r3}.
% 
% \begin{figure}[h]
% \includegraphics[width=200pt,angle=0]{NSEQS16.pdf}
% \caption{Composition of nuclear matter, $T=10$ MeV, $Y_p=0.1$, as function of the mass density. 
% the NSE without excited states (dotted) is compared to the QS in-medium approach with $A \le16$. 
% Continuum edge with self-energy shift according to DD2-RMF at $P=0$, Pauli blocking in scattering phase shifts by interpolation formula (\ref{eeffHe}),
% (\ref{eeffH}), 
% Pauli blocking of  $t, \alpha$ according $f,g$ of Tab. \ref{Tab:2c}.}
% \label{fig:NSEQS16}
% \end{figure}

Already at densities of  $10^{13} $ g/cm$^3$ the mass fraction of clustered, light $p$-shell nuclei has a maximum and start
to be blocked out. This happens before the neutron-rich isotopes like $^{10}$Be, $^{11}$Be become preferred Be isotopes
in the strongly asymmetric matter. We see that for the entire density region at $T,Y_p$ given above 
the mass fraction of nuclei with $Z > 2$ are below $2 \times 10^{-5}$ so that only a marginal change of the mass fractions of light nuclei ($A \le 4$)
is seen in comparison to Fig.  \ref{fig:NSEQS6}.

We focussed on the Pauli blocking effect as the main ingredient to determine the composition of nuclear matter near the saturation density.
For a more detailed investigation we should also consider other effects, in particular self-energy effects, see the Appendix \ref{App:SE}.
As discussed there, the single-nucleon self-energy shift should be reduced for heavier nuclei, and describing the $\bf p$ dependence in effective mass
approximation, additional changes of the composition at high densities are expected. For densities higher than the Mott density, i.e. $n_B> 0.03$ fm$^{-3}$,
the mass fraction of clustered, light $p$-shell nuclei is further reduced because we have only  continuum contributions which become small
for high densities, cf. Fig. \ref{fig:phase}. Correlations in the medium will further influence the in-medium modifications
as discussed in Ref. \cite{r3}. We expect that the composition in the subsaturation region $\rho > 10^{13.5} $ g/cm$^3$ is in general correctly described,
in particular the transition to a Fermi liquid of quasiparticles, but the detailed description of correlations as well as the stability against phase transitions 
remains open for future work.

\section{Conclusions}

Likewise the light  nuclei with mass number $A \le 4$, the exotic, light $p$-shell nuclei   ($4 \le A \le 16$) are strongly modified in warm dense matter.
Compared to the NSE, Pauli blocking leads to a reduction of the cluster abundances in nuclear matter and the dissolution of bound states at increasing densities.
The extension of the simple NSE to unstable nuclei like  $^4$H and $^5$He is problematic. The systematic treatment 
of continuum correlations using the generalized Beth-Uhlenbeck formulas gives a reduction of the mass fraction at increasing
temperatures and densities. The dominant appearance of neutron-rich unstable isotopes in asymmetric matter near the saturation density, 
as discussed, e.g., in Ref. \cite{Yudin} using improved versions of the NSE, is not supported. 

For fixed temperature and increasing density, correlations, in particular bound states, are formed in low-density nuclear matter according to the mass action law.
They are dissolved mainly owing to Pauli blocking near the saturation density  (Mott effect), 
and a single-nucleon quasiparticle approach to nuclear matter becomes applicable. 
%We find that at $T=10$ MeV, correlations in nuclear matter are significant only within a finite region of density around $n_{\rm sat}/10$. 
We obtained interpolation formula to describe the medium modifications
of clusters and correlations in dense nuclear matter in the temperature region $1\le T \le 20$ [MeV] and subsaturation densities.
Calculations of the composition of nuclear matter at $T=10$ MeV and $Y_p=0.1$ show, in addition to the light nuclei $d,\, t,\,h,\,\alpha$, 
a significant contribution of the neutron-rich clusters $^4$H and $^5$He in the density region around $n_{\rm sat}/10$, but at baryon number densities exceeding 0.03 fm$^{-3}$
these correlations are reduced by Pauli suppression.

Unbound nuclei like $^4$H and $^5$He are treated as correlations in the continuum. Using a generalized Beth-Uhlenbeck approach which is able to implement 
in-medium effects, the measured scattering phase shifts in the respective reaction channels have been used to give an estimate of the medium dependence of the phase shifts 
and to evaluate the partial densities of these components. For the bound $1p$ nuclei with mass number $6 \le A \le 16$, 
the intrinsic nucleon wave function of the $A$-nucleon cluster is an essential ingredient to evaluate the in-medium shift of the binding energy because of Pauli blocking.
Whereas for $10 \le A \le 16$ a shell model is applicable, the clustered nuclei with $6 \le A \le 9$ show a significant subcluster structure which has to be taken into account
to calculate the Pauli blocking shift. Simple fit formulas are given for the bound state energy  shifts  which can be used to evaluate the composition  
and related properties of warm and dense matter.

The quantum statistical approach is based on fundamental concepts such as the spectral function and the self-energy. 
Systematic improvements of our approach are possible considering further many-particle effects. In particular, at densities above 0.03 fm$^{-3}$
a detailed description of the self-energy as function of momentum and energy as well as correlations in the medium
may be a subject of future work to improve the description of correlations in nuclear matter near the saturation density.
Semi-empirical approaches such as the concept of excluded volume \cite{Hempel} or the generalized RMF \cite{TRB,Pais},
which are used presently to account for in-medium effects in the nuclear matter EoS, can profit to get inputs from a more systematic many-particle approach.

Astrophysical applications of the nuclear matter EoS demand the treatment of correlations in dense matter, see, e.g., \cite{Fischer,Fischerarx,Pais,Horowitz12}. 
For instance, the neutrino opacity of stellar matter is an important ingredient to describe supernova explosions.
The composition of nuclear matter and the formation of correlations determine the neutrino transport in hot and dense matter.
The large mass fractions of neutron-rich, unstable isotopes like $^4$He in stellar matter with low $Y_p$, which are predicted by NSE and related approaches  \cite{Yudin},
overestimate these correlations and are not appropriate for calculations of supernova and merger dynamics in the high-density region.

The understanding of few-body correlations in dense matter, in particular bound state formation, is an important ingredient  to describe heavy ion collisions. 
Light $p$ shell nuclei ($4 \le A \le 16$) are observed from HIC experiments, see Ref. \cite{Armstrong}. To explain the measured yields, 
different models can be used such as freeze-out of a fireball, coalescence models, or transport models like AMD or QMD simulations.
Compared to the simple NSE approximation for the freeze-out approach, we find a strong suppression of these yields  
at increasing densities because of Pauli blocking. Also for other approaches such as transport models \cite{Wolter}, 
in-medium effects should be taken into account to explain cluster formation in HIC.

The present work may be considered as a first step of a quantum statistical treatment of  light $1p$-shell clusters in nuclear matter. 
To describe the properties of hot and dense nuclear matter, in addition to improvements of  the approximations performed here,
we have also to extend the treatment of light clusters to heavier nuclei, see \cite{Debrecen,Gulminelli}. 
Light clusters, pasta phases, and phase transitions have to be considered  in core-collapse supernova matter \cite{collapse} and mergers. 
The investigation of few-nucleon correlations in dense matter is a basic prerequisite to understand matter under extreme conditions also
in nonequilibrium processes such as heavy ion collisions.

\begin{appendix}

\section{Solution of the in-medium wave equation for  $^6$Li}
\label{App:6Li}

The wave function of $^6$Li is taken as Gaussian,
\begin{equation}
 \psi_{^6{\rm Li}}^{\rm Gauss}(p_1,\dots,p_6)=\frac{1}{{\rm norm}_6}e^{-(p_1^2 +p_2^2+p_3^2+p_4^2+\beta_6 p_5^2+\beta_6 p_6^2)/B_6^2}p_{5,z}p_{6,z}
\delta_{{\bf p}_1+\dots + {\bf p}_6,0}\,.
\end{equation}
We adapt a Gaussian separable interaction
\begin{equation}
 V_{^6{\rm Li}}(12,1'2')=\lambda_6 \,e^{-\frac{({\bf p}_2-{\bf p}_1)^2}{4 \gamma_6^2}}e^{-\frac{({\bf p'}_2-{\bf p}'_1)^2}{4 \gamma_6^2}} \delta_{{\bf p}_1+{\bf p}_2,{\bf p}'_1+{\bf p}'_2}.
\end{equation}
The rms radius follows as (\ref{rms6beta})
\begin{equation}
\label{rmsLi}
{\rm rms}_{^6{\rm Li}}=\left[\frac{\beta_6}{6 B_6^2} \frac{21 + 160/\beta_6 + 382/\beta_6^2 + 688/\beta_6^3 + 288/\beta_6^4}{(1 + 
        2/\beta_6) (3 + 8/\beta_6 + 16/\beta_6^2)}  \right]^{1/2}.
\end{equation}
The intrinsic energy of the $^6$Li nucleus $ E^{(0)}_{^6{\rm Li}}={\rm KE}_{^6{\rm Li}}+{\rm PE}_{^6{\rm Li}}$ contains the kinetic and potential energy.
For the kinetic energy we find
\begin{equation}
{\rm KE}_{^6{\rm Li}}=\frac{\hbar^2}{2 m} \frac{B_6^2}{4 \beta_6}  \frac{21 + 160\beta_6 + 382\beta_6^2 + 688\beta_6^3 + 288\beta_6^4}{(1 + 
        2\beta_6) (3 + 8\beta_6 + 16\beta_6^2)}\,.
\end{equation}
The potential energy is
\begin{equation}
{\rm PE}_{^6{\rm Li}}=\lambda_6 (6 V^x_{12}V^y_{12}V^z_{12}+8 V^x_{15}V^y_{15}V^z_{15}+V^x_{56}V^y_{56}V^z_{56})
\end{equation}
with
\begin{equation}
 V^x_{12}=V^y_{12}=  V^z_{12} =\frac{B_6 \gamma_6^2}{\pi^{1/2}(B_6^2+4 \gamma_6^2)},
\end{equation}

\begin{eqnarray}
&& V^x_{15}=V^y_{15}=\frac{2 B_6 \gamma_6^2 (1+2 \beta_6)^{1/2} \beta_6^{1/2}}{\pi^{1/2}(B_6^2+  \gamma_6^2+ \beta_6 \gamma_6^2)^{1/2} (B_6^2 + 8 B_6^2 \beta_6 + 
   3 B_6^2 \beta_6^2 + 8 \beta_6 \gamma_6^2 + 16 \beta_6^2 \gamma_6^2)^{1/2}}\,,
\end{eqnarray}
$V^z_{15}$ is a lenghtly expression not given here, and

\begin{eqnarray}
&& V^x_{56}=V^y_{56}=\frac{B_6 \gamma_6^2 \beta_6^{1/2}}{\pi^{1/2}(B_6^2+2 \beta_6 \gamma_6^2)},\nonumber \\&& 
V^z_{56} =\frac{4 B_6 \gamma^2 \beta_6^{5/2}(3 B_6^4 - 4 B_6^2 \beta_6^2 + 4 B_6^2 \beta_6 \gamma_6^2 + 4 \gamma_6^4 + 8 \beta_6 \gamma_6^4 + 12 \beta_6^2 \gamma_6^4)}{\pi^{1/2}(3 + 8 \beta_6 + 16 \beta_6^2)(B_6^2 + 2 \beta_6 \gamma_6^2)^3}\,.
\end{eqnarray}

Within a variational approach, the optimum values for $B_6,\beta_6$ are obtained from the minimum of the intrinsic energy of the cluster.
The parameter $\lambda_6, \gamma_6$ are determined to reproduce the empirical values of the binding energy and the rms radius of $^6$Li, see Tab. \ref{Tab:1}. 
With $\lambda_6 = -964.5$ MeV fm$^3$ and $\gamma_6=1.16$ we find the optimum values 
for $B_6=1.0626$ fm$^{-1}$ and $\beta_6=3.6174$ so that $B_p=0.5587$ fm$^{-1}$. 
This result confirms the assumption that $B_s$
changes smoothly (see Tab. \ref{Tab:1}) whereas $\beta$ is rather large for  $^6$Li, and the result 
for $B_p$ agrees reasonably well with the estimated value shown in Tab. \ref{Tab:2}. 
Note that the ansatz for the wave function does not include deuteron-like clustering. Deuteron-like correlations are weak as shown by the low binding energy of $d$. A calculation including clustering is possible within the THSR ansatz. 

The potential energy is modified by the Pauli blocking effect. The evaluation of the shifts $\Delta E_{^6{\rm Li}}^{\rm Pauli,G}(T)=n_B\, \delta E_{^6{\rm Li}}^{\rm Pauli,G}(T)$ according Eq. (\ref{Awave})  gives the values shown in Tab. \ref{Tab:4} 
for different $T$. The value at $T=5$ MeV is also seen in Fig. \ref{fig:1}. It is higher than the shell model value. The large value of $\beta$ supports the extended  character of the $1p$ orbits leading to low-densities what favors the formation of clusters.
\begin{table}
\begin{center}
\hspace{0.5cm}
 \begin{tabular}{|c|c|}
\hline
$T$ &$\delta E_{^6{\rm Li}}^{\rm Pauli,G}(T)$  \\ 
{[MeV]} & [MeV fm$^3$]\\
\hline
1 & 3489.7 \\
2 & 3574.4  \\
3 & 3457.6  \\
4 & 3268.9  \\
5 & 3060.9  \\
6 & 2855.5  \\
7 & 2661.7  \\
8 & 2482.8  \\
9 & 2319.2  \\
10 & 2170.4  \\
12 & 1912.3  \\
14 & 1698.6  \\
16 & 1520.4   \\
18 & 1370.4  \\
20 & 1243.1  \\
\hline
 \end{tabular}
\caption{Pauli-blocking shifts $\Delta E_{^6{\rm Li}}^{\rm Pauli,G}(T,n_B)\approx n_B\, \delta E_{^6{\rm Li}}^{\rm Pauli,G}(T)$ for $^6$Li}
\label{Tab:4}
\end{center}
\end{table}

\begin{table}
\begin{center}
\hspace{0.5cm}
 \begin{tabular}{|c|c|c|c|c|c|c|c|c|c|c|}
\hline
$n_n$ [fm$^{-3}$] & $E_{\alpha n}^{\rm eff}$(2)&   $E_{\alpha n}^{\rm eff}$(4)&  $E_{\alpha n}^{\rm eff}$(6)&  $E_{\alpha n}^{\rm eff}$(8)&  $E_{\alpha n}^{\rm eff}(10)$& $E_{\alpha n}^{\rm eff}$(12)&   $E_{\alpha n}^{\rm eff}$(14)&  $E_{\alpha n}^{\rm eff}$(16)&  $E_{\alpha n}^{\rm eff}$(18)&  $E^{\rm eff}(20)$\\ 
\hline
0.0001 	& 1.525 	& 2.279 	& 3.317		& 4.639	& 6.236		& 8.091 		& 10.185 		& 12.499  	& 15.017 	& 17.723\\
0.001 	& 1.699 	& 2.481 	& 3.533		& 4.867	& 6.474 		& 8.339 		& 10.441 	 	& 12.764  	& 15.29 	& 18.003\\
0.01  	& 4.113 	 & 4.795	&5.931		& 7.372	& 9. 084 		& 11.047 		& 13.241	 	& 15.647  	& 18.246 	& 21.025\\
0.02  	& 7.196 	& 7.677 	& 8.859		& 10.42 	&12.27  		& 14.366 	 	& 16.681 	 	& 19.192 	& 21.882 	& 24.738\\
0.03  	& 10.3 	& 10.665 	& 11.904		& 13.603	& 15.614 		& 17.871 		& 20.33 	 	& 22.967  	& 25.764 	& 28.708\\
0.04 		& 13.344   & 13.669	& 14.997		& 16.86  	& 19.061 		& 21.503 		& 24.13 	 	& 26.912  	& 29.831 	& 32.875\\
0.05  	& 16.3        & 16.653 	& 18.112		& 20.173	& 22.595  		& 25.247 	 	& 28.062 	 	& 31.003  	& 34.054 	& 37.205\\
0.06  	& 19.167 	& 19.606 	&21.251		& 23.556	& 26.229   	& 29.113 	 	& 32.129 	 	& 35.238  	& 38.427 	& 41.689\\
0.07  	& 21.95 	& 22.547 	& 24.45		& 27.048	& 30.001 		& 33.131 	 	& 36.353 	 	& 39.632  	& 42.957 	& 46.327 \\
0.08  	& 24.668 	& 25.533 	& 27.787	 	& 30.727	& 33.975 		& 37.35	 	& 40.77 	 	& 44.208  	& 47.659 	& 51.129 \\
0.09  	& 27.389 	& 28.71	&31.414 		& 34.716 	& 38.245 		& 41.838 	 	& 45.429 	 	& 49.001  	& 52.557 	& 56.109\\
0.1   		& 30.426 	& 32.45 	& 35.613 		& 39.207 	& 42.938 		& 46.68 		& 50.386 	 	& 54.049  	& 57.678 	& 61.285 \\
\hline
$E_{\alpha n,0}(T)$  	&1.4157	 &  2.2051	&  3.2677		&  4.5979	& 6.1899 	&  8.033      & 10.113   &  12.413    & 14.918  & 17.613  \\
$E_{\alpha n,1}(T)$  	& 297.12	&  271.81	&  270.21		&  277.68	& 288.87 	&  301.18    & 313.25   &  324.39    & 334.35  & 343.04 \\
$E_{\alpha n,2}(T)$  	& -68.449	& 281.09 	&  494.93	 	&   649.07	 & 759.83  &  836.47   & 887.17  &  918.99    & 937.19  & 945.68  \\
\hline
 \end{tabular}
\caption{Second virial coefficient  (\ref{genBU}) and effective energies $ E_{\alpha n}^{\rm eff}(T)$ [MeV] according (\ref{efhe}), (\ref{eeffHe}). 
A separable potential (\ref{seppot}) was used with $\lambda = 670$ MeV fm$^3$, $\gamma = 1.791$ fm$^{-1}$, $T=2,4,\dots,18,20$ MeV.}
\label{Tab:P4}
\end{center}
\end{table}

\section{Separable potential model, $^5$He and $^4$H continuum correlations}
\label{Sec:separabel}

The solution of the in-medium wave equation to determine the medium corrections of the scattering phase shifts is convenient for 
a non-local, separable potentials. According to \cite{EST}, any potential can be represented as a sum of separable potentials. In nuclear physics, separable potentials are introduced, e.g., in Refs. \cite{Yama,Mongan}.

To reproduce the $\delta_{{\rm P}_{3/2}}(E)$ of the $\alpha-n$ scattering,
we use for the $l=1$ state
\begin{equation}
\label{seppot}
 V(p,p')=-\frac{\lambda}{\Omega_0}\,\frac{p}{(p^2/\gamma^2+1)^2}\,\frac{p'}{({p'}^2/\gamma^2+1)^2}.
\end{equation}
With
\begin{equation}
\label{Isep1}
 I(E)= \frac{\lambda}{2 \pi^2} \int_0^\infty dp\, \frac{p^2}{E-\hbar^2 p^2/(2m)} \frac{p^2}{(p^2/\gamma^2+1)^4}
\end{equation}
we have
\begin{equation}
\label{Isep2}
 I(E)= \frac{\lambda}{\pi}
\frac{\gamma^5 (-8 E^3-36  E^2 \gamma^2 \hbar^2/m+32 \sqrt{2}(-E  \hbar^2/m)^{3/2} \gamma^3+
18  E (\hbar^2/m)^2 \gamma^4+\gamma^6 (\hbar^2/m)^3)}{32(2 E + \gamma^2 \hbar^2/m)^4}.
\end{equation}

Bound states appear at $I(E)=-1$. The scattering phase shifts follow from
\begin{equation}
\label{Isep3}
 \delta_1(E)=-\arctan\left(\frac{{\rm Im} I(E)}{1+{\rm Re} I(E)}\right)\,.
\end{equation}
The empirical phase shifts of $^5$He are well reproduced with the parameter values $\lambda = 670$ MeV fm$^3$, 
$\gamma = 1.791$ fm$^{-1}$, see Sec. \ref{Sec:5He}. Parameter values for $^4$H are given in Sec. \ref{Sec:4H}.
%, see Fig. \ref{fig:7}. 
%However, the relation $1+{\rm Re} I(E^*) = 0$ gives 
%$E^* = 1.25975$ MeV so that $4/5 E^* = 1.0078$ MeV instead of 0.7356 MeV.

To include Pauli blocking effects, we have to consider in Eqs. (\ref{Isep1}),  (\ref{Isep3})
\begin{equation}
 I(E;T,\mu_n)= \frac{\lambda}{2 \pi^2} \int_0^\infty dp\, \frac{p^2}{E-\hbar^2 p^2/(2m)} \frac{p^2}{(p^2/\gamma^2+1)^4}[1-f_n(p;T,\mu_n)]\,.
\end{equation}
The neutron orbital is only blocked by the neutron background with density $n_n$. We assume an ideal fermion distribution with the chemical potential $\mu_n(T,n_n)$ according to 
\begin{equation}
 n_n = \frac{1}{\pi^2}\int_0^\infty dp \frac{p^2}{e^{(\hbar^2 p^2/2m-\mu_n)/T}+1}.
\end{equation}
The real part of $I(E;T,n_n)$ is given by the principal value integral, the imaginary part is
\begin{equation}
 {\rm Im}I(E;T,\mu_n)= \frac{\lambda}{2 \pi} \frac{m}{\hbar^2} \left(\frac{2 m E}{\hbar^2}\right)^{3/2}   
 \frac{1}{(2 E m/(\hbar^2\gamma^2)+1)^4}\left[1-\frac{1}{e^{E/T-\mu_n/T}+1}\right].
\end{equation}

We give some results for the virial coefficients calculated with in-medium scattering phase shifts in Tabs. \ref{Tab:P4},  \ref{Tab:CH4} and the corresponding effective energies which are used for the interpolations (\ref{Ean}), (\ref{Etn}).

\begin{table}
\begin{center}
\hspace{0.5cm}
 \begin{tabular}{|c|c|c|c|c|c|c|c|c|c|c|c|c|}
\hline
$n_n$ [fm$^{-3}$] &  $C^{\rm BU}(1)$  & $E^{\rm eff}(1)$&  $C^{\rm BU}(2)$  & $E^{\rm eff}(2)$&  
 $ C^{\rm BU}(3)$  & $E^{\rm eff}(3)$&$ C^{\rm BU}(5)$  & $E^{\rm eff}(5)$& $C^{\rm BU}(10)$  & $E^{\rm eff}(10)$ 
& $ C^{\rm gBU}(20)$  & $E^{\rm eff}(20)$ \\ 
\hline
0.0001 & 0.008651  & 4.7502 & 0.03475    & 6.7189  &  0.05832 & 8.5249  & 0.08764 & 12.183 & 0.1066  & 22.387 & 0.094462 & 47.191 \\
0.001  & 0.006673   & 5.0097 &  0.029987  & 7.0139 & 0.05206  & 8.8655 & 0.080605& 12.591  & 0.1014  & 22.882 & 0.091884& 47.744 \\
0.01   & 0.00008886 & 9.3283  & 0.0041078  & 10.989 & 0.01361  & 12.89 & 0.033803& 16.936 & 0.06141 & 27.901  & 0.06955  & 53.313 \\
0.02   & 6.236E-7   & 14.288 & 0.00036852 & 15.812  & 0.002727 & 17.713 & 0.012388& 21.955  & 0.0349  & 33.534  & 0.050911 & 59.554 \\ 
0.03   & 6.357E-9   & 18.874 & 0.00003619& 20.453 & 0.000564& 22.439 & 0.004588&26.921 & 0.01993& 39.155 &0.03721   & 65.826 \\ 
\hline
$ E_{\alpha n,0}(T)$ &  -  & 4.6196 &  -  &6.5808 &  -  & 8.4052  &  -  &  12.095 &  -  & 22.319 & - & 47.114 \\
 $E_{\alpha n,1}(T)$ &  -  & 477.15 &  -  & 459.68 &  -  & 464.96  &  -  & 492.79 &  -  & 560.79 & - & 622.89\\
\hline
\end{tabular}
\caption{$ C^{\rm BU}_{^4{\rm H}}(T)$  and  effective energies $ E_{tn}^{\rm eff}(T)$.
 A separable potential (\ref{seppot}) was used with $\lambda = 1144.9$ MeV fm$^3$, $\gamma =  1.326$ fm$^{-1}$, $T=1,2,3,5,10,20$ MeV.}
\label{Tab:CH4}
\end{center}
\end{table}

\section{Shifts and Mott densities for selected temperatures}
\label{Mott}

Within the shell model, the Pauli blocking shift is given by the contributions of occupied $1s$ and $1p$ orbitals according Eq. (\ref{delEP}). For symmetric matter we have
\begin{equation}
\label{delEPMott}
 \Delta E^{\rm Pauli}_{A,Z}(P=0;T,\mu_n,\mu_p) \approx n_B \, [4\, \delta E^{\rm Pauli}_{AZ,s}(T)+(A-4)\, \delta E^{\rm Pauli}_{AZ,p}(T)]= n_B \, \delta E^{\rm Pauli}_{A,Z}(P=0;T).
\end{equation}
For the light $p$-shell nuclei, both contributions are given in Tab.~\ref{Tab:3} at selected temperatures $T=5, 10, 15$, and 20 MeV.
The Mott density $n^{\rm Mott}_{A,Z}(T)=B_{A,Z}/ \delta E^{\rm Pauli}_{A,Z}(P=0;T)$ describes the baryon number density where at $P=0$ the bound state merge with the continuum and disappears. 
Values for $n^{\rm Mott}_{A,Z}(T)$ are also given in Tab. \ref{Tab:3}.
Note that above the Mott density $A$-nucleon correlations survive because bound states with large c.m.~momentum $\bf P$ can exist (the Pauli blocking becomes smaller, see Fig. \ref{fig:P}). In addition, $A$-nucleon continuum correlations remain. 

\begin{table}
\begin{center}
\hspace{0.5cm}
 \begin{tabular}{|c|c|c|c|c|c|c|c|c|c|c|c|c|c|}
\hline
$T$&= & 5 MeV &  5 MeV & 5 MeV &10 MeV &10 MeV &10 MeV &15 MeV &15 MeV &15 MeV &20 MeV &20 MeV &20 MeV \\
$A$  &  $Z$ &   $4 \delta E_{AZ,s}^{\rm Pauli} $ 
& $A' \delta E_{AZ,p}^{\rm Pauli} $ & $n_{A,Z}^{\rm Mott}$&   $4 \delta E_{AZ,s}^{\rm Pauli} $ 
& $A' \delta E_{AZ,p}^{\rm Pauli} $ & $n_{A,Z}^{\rm Mott}$&   $4 \delta E_{AZ,s}^{\rm Pauli} $ 
& $A' \delta E_{AZ,p}^{\rm Pauli} $ & $n_{A,Z}^{\rm Mott}$&   $4 \delta E_{AZ,s}^{\rm Pauli} $ 
& $A' \delta E_{AZ,p}^{\rm Pauli} $ & $n_{A,Z}^{\rm Mott}$\\
\hline
4& 2 & 2389.7 & - 		&0.01184& 1691.5 & -		& 0.01673& 1277.3 & -		& 0.02216& 1008.2 & -		 & 0.02807\\
 6 & 3 & 2596.8& 226.8 	&0.01133&1777.7&175.5 	& 0.01638& 1313.5 & 132.5	& 0.02212& 1021.1 & 103.2 & 0.02846 \\
 7 & 3 & 2695.5 &  422.7 & 0.01258 & 1821.0 &  374.4 	& 0.01788  & 1334.2 &  304.0 	& 0.02395 & 1031.2 &  247.8 & 0.03068 \\
 7 & 4 & 2695.5 &  378.1 & 0.01223 & 1821.0 &  307.8 	& 0.01766 & 1334.2 &  238.7 	& 0.02391 & 1031.2 &  188.9 & 0.03082 \\
 9 & 4 & 2888.5 &  796.3 & 0.01578& 1908.6 &  709.1 	& 0.02222 & 1379.3 &  577.5 	& 0.02972 & 1056.1 &  471.5 & 0.03807 \\
10 & 4 & 2976.8 & 1240.6 & 0.01541 & 1948.2 & 1236.8 & 0.02040 & 1399.7 & 1076.1 	& 0.02624 & 1067.5 &  917.3 & 0.03275 \\
11 & 4 & 3074.9 & 1453.6 & 0.01446 & 1995.1 & 1401.2 & 0.019281 & 1425.9 & 1194.5 & 0.02499 & 1083.6 & 1004.5 & 0.03136 \\
10 & 5 & 2976.8 & 1172.5 & 0.01561 & 1948.2 & 1133.7 & 0.02101& 1399.7 &  968.3 	& 0.02734 & 1067.4 &  815.4 & 0.03439 \\
11 & 5 & 3074.9 & 1520.9 & 0.01658 & 1995.1 & 1500.5 & 0.02180 & 1425.9 & 1297.1 & 0.02799 & 1083.6 & 1100.9 & 0.03469 \\
12 & 6 & 3161.4 & 1792.7 & 0.01860 & 2034.5 & 1737.1 & 0.02443 & 1446.9 & 1485.5 & 0.03143 & 1095.9 & 1251.7 & 0.03927 \\
13 & 6 & 3257.5 & 2198.8 & 0.01779 & 2081.2 & 2152.2 & 0.02294 & 1473.7 & 1851.4 & 0.02921 & 1113.0 & 1566.2 & 0.03624 \\
14 & 6 & 3342.0 & 2537.8 & 0.01791 & 2119.8 & 2456.5 & 0.02301 & 1494.6 & 2099.0 & 0.02930 & 1125.7 & 1767.7 & 0.03639 \\
14 & 7 & 3342.0 & 2419.4 & 0.01817 & 2119.8 & 2294.1 & 0.02371 & 1494.6 & 1936.4 & 0.03051 & 1125.7 & 1617.6 & 0.03815 \\
15 & 7 & 3437.2 & 2738.7 & 0.01870 & 2165.6 & 2562.8 & 0.02442 & 1520.9 & 2146.4 & 0.03149  & 1142.5 & 1783.9 & 0.03946 \\
16 & 8 & 3524.6 & 2945.7 & 0.01972 & 2206.9 & 2678.4 & 0.02612  & 1544.3 & 2206.2 & 0.03403 & 1157.3 & 1813.7 & 0.04295 \\
\hline
 \end{tabular}
\caption{Temperature-dependent shifts $\delta E_{AZ,s}^{\rm Pauli}(T) $ and $\delta E_{AZ,p}^{\rm Pauli}(T) $ 
of $1s,\,1p$ nuclei, $A'=A-4$. 
The corresponding Mott densities $n_{A,Z}^{\rm Mott}(T)$ are also given. Units: MeV, fm.}
\label{Tab:3}
\end{center}
\end{table}

\section{Single-nucleon self-energy shifts}
\label{App:SE}

Single-particle excitations in nuclear systems are described by the single-nucleon spectral function
$A_\tau({\bf p}, \omega)$,
which is determined by the dynamical self-energy $\Sigma_\tau({\bf p}, \omega)$.
In the quasiparticle approach the  spectral function is approximated by a $\delta$-like single-particle contribution and a background. The quasiparticle dispersion relation 
$E_\tau^{\rm qu}({\bf p})= \hbar^2 p^2/(2 m_\tau)+\Delta E^{\rm SE}_\tau({\bf p})$ contains
the self-energy shift $\Delta E^{\rm SE}_\tau({\bf p})=\Sigma_\tau[{\bf p},\omega =E_\tau^{\rm qu}({\bf p} )]$.

The concept of quasiparticle excitation proved to be successful at low densities as well as 
high densities (Fermi liquid). A criterion is that further correlations which determine the background 
of the spectral function are not significant. The quasiparticle shift can be related to empirical data,
we use here the DD2-RMF approximation \cite{Typel}. For instance, at $T=0$ the quasi-particle shift in the low-density limit amounts (units MeV, fm)
\begin{eqnarray}
\label{RMFnB}
&&\Delta E^{\rm SE, RMF}_\tau(T=0,n_B,Y_p) \approx  \\ &&
[-1058.4+490.15\,{\rm sign}_\tau (1-2 Y_p)-1.659 (1-2 Y_p)^2-0.00761 \,{\rm sign}_\tau (1-2 Y_p)^3
-0.2668 (1 - 2 Y_p)^4] n_B + {\cal O}(n_B^2),\nonumber
\end{eqnarray}
with ${\rm sign}_\tau=1$ for $\tau =n$ and ${\rm sign}_\tau=-1$ for $\tau =p$.

The microscopic approach to the self-energy can be performed using the method of Green's functions 
and diagram representations, but needs also an expression for the interaction potential. 
In lowest order of interaction, we obtain the Hartree-Fock approximation 
\begin{equation}
 \Delta E^{\rm SE, HF}_{\tau_1}(1) = \sum_2 [V(12,12)-V(12,21)] f(2).
\end{equation}
For instance, the Hartree shift of the simple Yukawa potential 
\begin{equation}
V^{\rm Yukawa}(r)=-\lambda \exp[-r/R_\pi]/r
\end{equation} 
with $R_\pi = 1.4$ fm reproduces (\ref{RMFnB}) for symmetric matter ($Y_p=1/2$) if the parameter value 
$\lambda = 42.97$ MeV fm is chosen.

There exist more realistic nucleon-nucleon interactions and higher order diagram approximations 
such as the Brueckner-Hartree-Fock approximation considering ladder sums for the self-energy, 
see also \cite{SRS}.
Correlations in the medium may be taken into account leading to the cluster mean-field 
approximation \cite{r2}. The consideration of higher order diagrams allows to introduce the two-particle 
distribution function as known from the average potential energy in classical statistics.

For nuclei, the effective interaction with free nucleons can be modeled by an optical potential. 
A recent version for $A \le 13$ has been given in Ref. \cite{Weppner}. 
It can be approximated by the folding integral of the Yukawa interaction with the density distribution 
of nucleons in the core nucleus. We will not present here details of such model calculations.

Whereas the interaction outside the nucleus is reasonably described by a Yukawa-like potential, 
see also the M3Y potential \cite{M3YReview},
slow free nucleons can only hardly enter the nucleus because of the Pauli principle. This is 
a higher order effect, the core nucleon is part of the cluster which determines the Pauli-forbidden 
region in the phase space.
In higher orders of perturbation theory we have to include diagrams leading to
 the pair distribution function $g(r )$, 
as also seen in the cluster mean-field approximation \cite{r3}. 
The pair distribution function becomes small at the surface of the core nucleus, radius
$R_A = (4 \pi n_{\rm sat}/3)^{-1/3} A^{1/3} = 1.17  A^{1/3}$ fm.
As a consequence, the mean field given by the average potential is also reduced if short distances 
between the cluster nucleon and the free environmental nucleon are avoided.
This effect should be taken into account if larger nuclei in matter are considered. 
A more sophisticated evaluation, considering higher order diagrams to describe the antisymmetrization 
of the outside free nucleon with respect to all nucleons bound in the cluster, 
may be a topic of future investigations if heavy clusters in matter are considered.

To give an estimate of the reduction of the nucleon pair distribution inside the nucleus, 
we cut the Hartree term at the surface of the nucleus,
\begin{equation}
 \Delta E^{\rm SE, cut} =4 \pi \int_{R_A}^\infty dr \, r^2 \,V^{\rm Yukawa}(r ).
\end{equation}
Compared to the value at $R_A=0$, we have the reduction %0.78 for $A=4$, 
0.68 for $A=6$, 
and 0.48 for $A=16$.
We conclude that the single-nucleon self-energy shift is reduced by a factor of about 1/2 
in the region $6 \le A \le 16$. 

This effect will not change the general feature of the formation and dissolution of clusters 
when the baryon density rises to the saturation density. 
It is not of relevance for the composition at low densities, 
but modifies the composition near $n_{\rm sat}$. 
Note that further effects are obtained from the effective mass corrections, see \cite{r1}. 
With the empirical value for the effective mass given there, the binding energy of 
the cluster is slightly reduced.

\end{appendix}


\begin{thebibliography}{10}
\footnotesize
\renewcommand{\baselinestretch}{0.5}


\bibitem{RS}
 P. Ring and P. Schuck, {\it The Nuclear Many-Body Problem} (Springer, Berlin 1980).

\bibitem{THSR}
A. Tohsaki, H. Horiuchi, P. Schuck, and G. R\"opke,  Phys. Rev. Lett. {\bf 87}, 192501 (2001).

\bibitem{THSR2}
A. Tohsaki {\it et al.}, Rev. Mod. Phys. {\bf 89}, 011002 (2017).

\bibitem{Xu}
Shuo Yang {\it et al.}, Phys. Rev. C {\bf 101}, 024316 (2020).

\bibitem{Armstrong}
T. A. Armstrong {\it et al.}, Phys.Rev. C {\bf 65}, 014906 (2001).

\bibitem{Ganil}
H. Pais {\it et al.}, arXiv:1911.10849 (2019).

\bibitem{RMS82}
G.\ R\"opke,  L.\ M\"unchow, and H.\ Schulz,
  Nucl. Phys. A {\bf 379}, 536 (1982);
Phys. Lett. {\bf B 110}, 21 (1982). 

\bibitem{NSE}
A.~S.~Botvina and I.~N.~Mishustin,
Nucl.\ Phys.\ A {\bf 843}, 98 (2010).\\
N. Buyukcizmeci  {\it et al.},
%, A. Botvina, I. Mishustin, R. Ogul, M. Hempel, J. Schaffner-Bielich, F.-K. Thielemann, S.
%Furusawa, K. Sumiyoshi, S. Yamada, and H. Suzuki,
Nucl.Phys. A {\bf 907}, 13 (2013).

\bibitem{BU}
E. Beth and G. E. Uhlenbeck, Physica {\bf 4}, 915 (1937).

\bibitem{HS}
C. J. Horowitz and A. Schwenk, Nucl. Phys. A {\bf 776}, 55 (2006).

\bibitem{Wolter}
Jun Xu {\it et al.}, Phys. Rev. C {\bf 93}, 044609 (2016).

\bibitem{Fischer}
 T. Fischer {\it et al.},  ApJS {\bf 194}, 39 (2011).
 
\bibitem{Fischerarx}
T. Fischer {\it et al.},  arxiv:2003.00972 (2020).
 
\bibitem{SRS}
M.\ Schmidt {\it et al.},
Ann. Phys. (N.Y.) {\bf 202}, 57 (1990).

\bibitem{r3}
G.~R\"opke, Phys. Rev. C {\bf 92}, 054001 (2015).

 \bibitem{Debrecen} 
G.~R\"opke, J. Phys.: Conf. Series {\bf 436}, 012070 (2013).

\bibitem{Yudin}
A. V. Yudin, M. Hempel, S. I. Blinnikov, D. K. Nadyozhin, and I. V. Panov, 
Monthly Notices of the Royal Astronomical Society (MNRAS), {\bf 483}, 5426 (2019).

\bibitem{Hempel}
M. Hempel and J. Schaffner-Bielich, 
Nucl. Phys. A {\bf 837}, 210 (2010).

\bibitem{Pais}
H. Pais, F. Gulminelli, C. Providencia, and G. R\"opke
Phys. Rev. C {\bf 99}, 055806 (2019).

\bibitem{TRB}
S. Typel {\it et al.}, Phys. Rev. C {\bf 81}, 015803 (2010).


\bibitem{AudiWapstra}
G. Audi {\it et al.}, Nuclear Physics A {\bf 729}, 337 (2003);\\
Meng Wang, G. Audi {\it et al.},  Chinese Physics C {\bf 41},  030003 (2017).

\bibitem{nuclei}
NuDat/Chart of Nuclides (http://www.nndc.bnl.gov);\\
https://www-nds.iaea.org/relnsd/vcharthtml/VChartHTML.html.

\bibitem{clustervirial} 
G.~R\"opke {\it et al.}, Nucl. Phys. A {\bf 897}, 70 (2013).
	
\bibitem{VT}
M. D. Voskresenskaya and S. Typel, Nucl. Phys. A {\bf 887}, 42 (2012).

\bibitem{Typel}
	S.~Typel,
	Phys. Rev. C {\bf 71}, 064301 (2005);\\
	S.~Typel and H.~H. Wolter,
	Nucl. Phys. A {\bf 656}, 331 (1999).

\bibitem{r1}
G.~R\"opke, Phys. Rev. C {\bf 79}, 014002 (2009).

\bibitem{r2}
G.~R\"opke, 
Nucl. Phys. A {\bf 867}, 66 (2011).

\bibitem{WoodsS}
N. Schwierz, I. Wiedenhover, and A. Volya,  arxiv:0709.3525

\bibitem{Yama}
Y. Yamaguchi,  Phys. Rev. {\bf 95}, 1628 (1954).

\bibitem{Mongan}
T. R. Mongan, Phys. Rev. {\bf 175}, 1260 (1968); {\bf 178}, 1597 (1969); {\bf 180}, 1514 (1969).

\bibitem{EST}
D. J. Ernst, C. M. Shakin, and R. M. Thaler, Phys. Rev. C {\bf 8}, 46, 2056 (1973).

\bibitem{solar}
 M. Asplund {\it et al.}, Annual Review of Astronomy and Astrophysics {\bf 47}, 481 (2009).
 
\bibitem{rmsradii}
I. Angeli and K.P. Marinova, Atomic  Data and Nuclear Data Tables {\bf 99}, 69 (2013).

\bibitem{Lyu}
Mengjiao Lyu {\it et al.},
%, Zhongzhou Ren, Bo Zhou, Yasuro Funaki, Hisashi Horiuchi, Gerd R{\"o}pke, Peter Schuck, Akihiro Tohsaki, Chang Xu, and Taiichi Yamada,
Phys. Rev. C {\bf 91}, 014313 (2015).

\bibitem{Zhao}
Qing Zhao {\it et al.},
%, Zhongzhou Ren, Mengjiao Lyu, Hisashi Horiuchi, Yoshiko Kanada-En'yo, Yasuro Funaki, Gerd R{\"o}pke, Peter Schuck, Akihiro Tohsaki, Chang Xu, Taiichi Yamada, and Bo Zhou,
Phys. Rev. C {\bf 100}, 014306 (2019).

\bibitem{Kanada}
Y. Kanada-En'yo, H. Horiuchi, Prog. Theor. Phys. {\bf 142}, 205 (2001).

%%%%%%%%%%%%%%%

 \bibitem{Wild1977}
 K. Wildermuth and Y. C. Tang, {\it A Unified Theory of the Nucleus} (Vieweg, Braunschweig, Germany, 1977). 

\bibitem{Saito1977}
S. Saito, Prog. Theor. Phys. Suppl. {\bf 62}, 11 (1977).

\bibitem{Hori1977}
H. Horiuchi, Prog. Theor. Phys. Suppl. {\bf 62}, 90 (1977).

\bibitem{Hori2012}
H. Horiuchi, K. Ikeda, and K. Kato, Prog. Theor. Phys. Suppl. {\bf 192}, 1 (2012).




\bibitem{Wiringa}
R. B. Wiringa, S. C. Pieper, J. Carlson, and V. R. Pandharipande, 
Phys. Rev. C {\bf 62}, 014001 (2000).

\bibitem{Egorov}
M.V. Egorov, Nucl. Phys. A {\bf 987}, 247  (2019).

\bibitem{Huang}
K. Huang, {\it Statistical Mechanics}, second ed., (Wiley, New York, 1987).

\bibitem{Hoop}
B. Hoop and H. H. Barschall, Nucl. Phys. {\bf 83}, 65 (1966).

\bibitem{Arndt}
R. A. Arndt and L. D. Roper, Phys. Rev. C {\bf 1}, 903 (1970).


\bibitem{Seagrave}
J. D. Seagrave {\it et al.}, Ann Phys. (N.Y.) {\bf 74}, 250 (1972).

\bibitem{LeMere}
M. LeMere  {\it et al.}, Phys. Rev. C {\bf 12}, 1140 (1975).

\bibitem{Shen}
P. N. Shen {\it et al.},  Phys. Rev. C {\bf 33}, 1214 (1986).

\bibitem{Reichstein}
I. Reichstein, D. R. Thompson, and Y. C. Tang,  Phys. Rev. C {\bf 3}, 2139 (1971).

\bibitem{Connor}
E. O'Connor {\it et al.}, Phys. Rev. C {\bf 75}, 055803 (2007).

\bibitem{Horowitz12}
C. J. Horowitz {\it et al.}, Phys. Rev. C {\bf 86}, 065806 (2012).

\bibitem{Weppner}
S. P. Weppner, J. Phys. G: Nucl. Part. Phys. {\bf 45}, 095102 (2018).

\bibitem{M3YReview} 
G. R. Satchler and  W. G. Love, Phys. Rep. {\bf 55}, 183 (1979).

\bibitem{Gulminelli}
%C. Ducoin, P. Chomaz, and F. Gulminelli, Nucl. Phys. {\bf A781},
%407 (2007).\\
 F. Gulminelli and Ad. R. Raduta, Phys. Rev. C {\bf 92}, 055803 (2015);\\
Ad. R. Raduta and F. Gulminelli, Phys. Rev. C {\bf 82}, 065801
(2010).\\
H. Pais  {\it et al.},
%, F. Gulminelli, C. Providncia, and G. Ršpke, 
Phys. Rev. C {\bf 97}, 045805 (2018).

\bibitem{collapse}
H. Pais {\it et al.}, Phys. Rev. C {\bf 91}, 055801 (2015)




%
%\end{thebibliography}

%\begin{thebibliography}{10}
%\footnotesize
%\renewcommand{\baselinestretch}{0.5}


%%%%%%%%%%%%%%%%%%%%%%%%%%%%%%%%%%%%
%%%%%%     general references    %%%%%%%%%%
%%%%%%%%%%%%%%%%%%%%%%%%%%%%%%%%%%%%%%%%%


\end{thebibliography}
\end{document}